\documentstyle[epsf,rotate,psfig]{mn}
\def\gsim{~\rlap{$>$}{\lower 1.0ex\hbox{$\sim$}}}
\def\etal{{\em et al. }}
\def\ia{\'{\i}}

\title{Ages and Metallicities of Globular Clusters in NGC~4472}
 
\author[M.A.~Beasley et al.]
  {M.A.~Beasley,$^1$\thanks{email: m.a.beasley@durham.ac.uk} R.M.~Sharples,$^1$ T.J.~Bridges,$^2$ D.A.Hanes,$^3$ S.E.~Zepf,$^4$ K.M.~Ashman,$^5$ 
\newauthor
 D.~Geisler,$^6$ \\
  $^1$Department of Physics, University of Durham, Durham DH1 3LE, UK\\
  $^2$Anglo-Australian Observatory, P.O. Box 296, Epping, NSW, 2121, Australia\\
  $^3$Department of Physics, Queen's University, Kingston, ON K7L 3N6, Canada\\
  $^4$Department of Astronomy, Yale University, New Haven, CT 06520\\
  $^5$Department of Physics and Astronomy, University of Kansas, Lawrence
KS 66045\\
  $^6$Departamento de F\ia sica, Grupo de Astronom\ia a, Universidad de Concepci$\acute{o}$n, Casilla 160-C, Concepci$\acute{o}$n, Chile}

\date{}
\pagerange{\pageref{firstpage}--\pageref{lastpage}}
\pubyear{2000}

\def\LaTeX{L\kern-.36em\raise.3ex\hbox{a}\kern-.15em
    T\kern-.1667em\lower.7ex\hbox{E}\kern-.125emX}

\begin{document}

\label{firstpage}

\maketitle

\begin{abstract}

We have derived ages and metallicities from co-added spectra of 
131 globular clusters associated with the giant elliptical galaxy 
NGC~4472.
Based upon a calibration with Galactic globular clusters, we find that 
our sample of globular clusters in NGC~4472 span a metallicity range of 
approximately -1.6 $\leq$ [Fe/H] $\leq$ 0 dex.
There is evidence of a radial metallicity gradient in the globular cluster
system which is steeper than that seen in the underlying 
starlight. 
Determination of the \emph{absolute} ages of the globular clusters is 
uncertain, but formally, the metal-poor population of globular clusters has
an age of 14.5 $\pm$ 4 Gyr and the metal-rich population is 
13.8 $\pm$ 6 Gyr old.
Monte Carlo simulations indicate that the globular cluster populations 
present in these data are older than 6 Gyr at the 95\% confidence 
level. We find that within the uncertainties, the globular clusters are 
old and coeval, implying that the bimodality seen in the 
broadband colours primarily reflects metallicity and not age differences.

\end{abstract}


\section{Introduction}

It has become clear in recent years that subpopulations in extragalactic
globular cluster systems (GCS) are not uncommon. Indeed, since their first 
detection by Zepf \& Ashman (1993) using colour data from Couture, Harris
\& Allwright (1991) and Harris \etal (1992), the study
of colour bimodality in GCSs has become a growth industry 
(e.g. Neilsen \& Tsvetanov 1999; Gebhardt \& Kissler-Patig 1999).
It is now estimated that \emph{at least} half of all early-type systems
have bimodal GCSs (Gebhardt \& Kissler-Patig 1999).
What is not so clear is the origin of these subpopulations, and their 
implications for the formation of their host galaxies.
The apparent ubiquity of globular clusters in galaxies, and the homogeneity 
of their GCSs 
naturally leads one to conclude that they are intimately associated with, 
if not the epoch of galaxy formation, at least its major modes of 
star formation.

NGC~4472, the brightest elliptical galaxy in Virgo, is an excellent 
candidate for such GCS research and has consequently been the subject of 
extensive photometric studies (e.g. Cohen 1988; Couture \etal 1991; 
Geisler, Lee \& Kim 1996; Lee, Kim \& Geisler 1998). 
This galaxy (Hubble type E2) plays host to some
6300 $\pm$ 1900 globular clusters (Geisler \etal 1996) which have been 
shown to possess a bimodal colour distribution 
(Zepf \& Ashman 1993; Geisler \etal 1996). 
Adopting a distance modulus of $m-M$ = 31.0, gives an absolute 
magnitude of $M_{V}= $ -22.6 for this galaxy, and a specific frequency of 
S$_{\rm{N}}=$ 5.6 $\pm$ 1.7 (Geisler \etal 1996).

To first order, the GCS colour distributions are thought to reflect
the distribution of metallicity in the globular cluster populations
(Ashman \& Zepf 1998).
Since globular clusters are generally thought to be old, 
their relative ages will not significantly affect broadband colours 
owing to the well known age-metallicity degeneracy (Faber 1972; O'Connell
 1976; Worthey 1994).
The problem of breaking this degeneracy, and extracting relative age 
information about globular cluster subpopulations has lead to several
avenues of investigation.
One technique has been to compare broadband colours (which are primarily 
metallicity-sensitive) with magnitudes (which are largely age sensitive, 
assuming a given mass-to-light ratio).
Puzia \etal (1999) have investigated the ages of the globular cluster
subpopulations in NGC~4472, using HST $V$ and $I$ band photometry to determine
both mean colours and turn-over magnitudes. 
By assuming that the mean mass functions of the two subpopulations 
are similar, they find that 
the blue and red globular clusters are coeval within their 
uncertainties (which are $\pm$ 3 Gyr). They conclude that possibility of
one population being half the age of the other is excluded at the 
99\% confidence level. However, Lee \& Kim (2000) have also analysed
HST $V$ and $I$ data for NGC~4472, using three datasets in common with the  
Puzia \etal (1999) study, and arrive at a different conclusion.
 They indicate that the red globular clusters are several Gyr \emph{younger} 
than the blue globulars. 
This difference from the Puzia \etal (1999) study they attribute
to a combination of a $\sim$ 30\% larger sample
size, greater areal coverage and increased completeness. 
A similar result has been found by Kundu \etal (1999) for globular clusters
 in the inner regions of NGC~4486 (M87). 
They find that the red globular clusters are 3--6 Gyr 
younger than their blue counterparts, with the
magnitude of the age difference dependent upon the stellar evolutionary 
models they adopt.

An alternative method is via spectroscopy, which can provide a way of 
disentangling age from metallicity through the careful selection of 
absorption-line indices. This approach circumvents the requirement of
assumptions of a universal mass distribution for GCs. 
Metallicities may be directly measured from metal-sensitive features, and  
ages may be found from Balmer indices, which effectively measure the
temperature of the main-sequence turnoff in globular clusters.
In a previous paper, Sharples \etal (1998) obtained spectra for 47 globular 
clusters associated with NGC~4472, and including data from the study of 
Mould \etal (1990), 
found evidence for kinematical differences between the blue and red cluster 
populations. Another 100 spectra for \emph{bona fide} globular clusters 
associated with this galaxy have now been obtained with the Multi--Object
Spectrograph (MOS) at the 3.6-m Canada-France-Hawaii telescope (CFHT), 
creating a total dataset of 131 confirmed globular clusters. 
The kinematic analysis of these data is presented in Zepf \etal (in 
preparation), in this paper we concentrate on the abundances and ages 
of the NGC~4472 globular clusters.

The structure of this paper is as follows: in $\S 2$ we briefly overview 
previous spectroscopy for extragalactic GCSs in luminous ellipticals
and in $\S 3$ the observations and data reduction are described. 
This is followed in $\S 4$ by the procedure for measuring line-strengths of
the globular clusters. In $\S 5$ we present both the qualitative and
quantitative results from this work. We discuss the implications of
our findings in $\S 6$ and summarize our work in $\S 7$.

\section{Previous Spectroscopy of GCS in Luminous Elliptical Galaxies}

The faintness of extragalactic GCS and the demands of spectroscopic studies  
upon instrumentation and telescope time have led to a situation where only a 
few line-strength measurements have been successfully undertaken for 
globular clusters outside the Local Group.
Hanes \& Brodie (1986) derived a mean metallicity of the M87 
globular cluster system of [Fe/H] = -0.5 $\pm$ 0.4 dex utilising low S/N
 and moderately low resolution integrated spectra. 
This is consistent with a value of [Fe/H]= -0.89 $\pm$ 0.16 dex found by 
Brodie \& 
Huchra (1991) using the MMT spectrograph at $\sim$ 9.0 \AA\ resolution. 
Mould \etal (1990) use the median Mg $b$ index  
and calibrations of Burstein \etal (1984) in order to obtain a mean 
metallicity of -1.0 $\pm$ 0.2 dex for M87 and -0.8 $\pm$ 0.3 dex for NGC~4472. 
More recently, Kissler-Patig \etal (1998) used the Keck telescope to 
obtain 5.6 \AA\ resolution 
spectra of 21 globular clusters belonging to NGC~1399, the central cD galaxy 
of the Fornax cluster. 
They derive a mean [Fe/H] of 
-0.8 dex and find that the elemental abundances of the majority of the 
globular clusters do not differ from those observed in the Milky Way and M31. 
However, two of their globular clusters stand out in that their metallic line indices 
are comparable to the diffuse stellar light of the host galaxy, 
perhaps implying that they may have formed from nearly solar 
metallicity gas in a different formation process from the bulk 
of the globular clusters (such as may be attributed to a merger event). 
They also exhibit Balmer line indices (H$\beta$, H$\gamma$) incompatible 
with \emph{any} age-metallicity combination of existing stellar population
 models. Cohen, Blakeslee \& Ryzhov (1998) analyse a larger sample of 150 
globular clusters in M87, and show that 
the cluster system spans a wide range of metallicity, -2.2 $<$ [Fe/H] $<$ 0.1, 
dex with a mean [Fe/H] of -0.95 dex. 
They find marginal evidence for a bimodal as opposed to unimodal metallicity 
distribution at the 86 \% significance level, with 'peaks' at 
-1.3 and -0.7 dex, however this may be largely due to their object
selection biases.
Through the measurement of H$\beta$ and H$\alpha$ 
indices, they obtain a median age for the M87 GCS of 
13~$\pm$ 2 Gyr, similar to that of the Milky Way globular clusters.
 
\section{Observations and Data Reduction}

The data constituting our combined sample was obtained from two separate 
observing runs. 
Spectroscopic observations of 79 cluster candidates were taken with 
the Low-Dispersion Survey Spectrograph on the 4.2-m William Herschel 
Telescope (WHT) in April, 1994 
(see Sharples \etal 1998 for further details of these observations and the 
data reduction).
This has now been supplemented by spectra of 171 cluster candidates obtained 
using MOS on the CFHT in May, 1998.
Object selection in both studies was based on 
Washington photometry of NGC~4472 globular cluster candidates published
by Geisler \etal (1996). 
Five multislit masks were prepared using the MOS software 
 provided at the CFHT, though only four were actually used.
Slits were cut using the \textsc{lama} machine, with typically 
40 $\sim$ 45 objects per mask.
One of the masks was centred on the nucleus of NGC~4472,
with the other four being displaced by $\sim$ 5\farcm5 into the
NE, NW, SE, and SW quadrants to obtain the best
spatial coverage of the cluster system. Highest priority was 
given to candidates in the magnitude range $19.5< V <21.5$,
although candidates were selected down to $V$=22.5. Table~\ref{tab:masks}
gives the details of the masks, listing the number of candidates per mask, 
along with the number of spectroscopically confirmed globular clusters, 
background galaxies and foreground stars. A number of spectra yielded no
reliable identification, due to insufficient signal-to-noise (S/N)
and/or problematical sky-subtraction. 

\begin{table*}
\caption{\small{Details of the mask setup and spectroscopic completeness for
the CFHT observations.}}
\label{tab:masks}
\begin{tabular}{ccccccc}
\hline 
   Mask & RA & DEC & Objects/ & Confirmed & Background & Foreground\\
\# & (1950.0) & (1950.0) & Mask & Globulars & Galaxies & Stars\\ 
\hline


Mask1 & 12 27 15.14 & 08 16 37.4 & unused & unused & unused & unused \\
Mask2 & 12 27 01.69 & 08 12 36.3 & 47 & 22 & 4 & 2 \\
Mask3 & 12 27 01.69 & 08 20 38.3 & 42 & 26 & 3 & 0 \\
Mask4 & 12 27 27.06 & 08 20 38.4 & 42 & 31 & 3 & 2 \\
Mask5 & 12 27 27.06 & 08 12 36.4 & 45 & 21 & 3 & 3 \\

\hline

\end{tabular}
\end{table*}

The B600 grism with dispersion 2.24~\AA\ pix$^{-1}$ was used, producing 
spectra with an instrumental resolution of $\sim$ 5.5~\AA\  
(330 kms$^{-1}$) and a useful spectral range of 3800 -- 6500~\AA. 
The detector was a STIS 2048$^{2}$ chip, with readout noise 9.3 e$^{-}$, 
noteworthy for its excellent performance in the blue 
(quantum efficiency $\sim$ 82 \% at 4000~\AA).
Flat field and bias frames were taken at the beginning and end of each night, 
and the spectra were wavelength calibrated using frequent mercury arcs taken 
before and after the programme object frames. 
Table~\ref{tab:obslog} lists the observational details for the CFHT run.

\begin{table}
\begin{center}
\caption{\small{Observing log for CFHT run.}}
\label{tab:obslog}
\begin{tabular}{ll}
\hline 

Dates & May 16-19 1998\\
Telescope/Instrument & 3.58m CFHT/MOS\\
Detector & 2048$^{2}$ STIS-2\\
Dispersion (Resolution) &  2.24 \AA\ pix$^{-1}$ (5.5 \AA\ FWHM)\\
Wavelength Coverage & 3800 - 6500 \AA \\
Seeing & $< 1''$\\
Mean Airmass & 1.1\\
Exposure times (NGC~4472) & 6000s per mask\\

\hline

\end{tabular}
\end{center}
\end{table}

For velocity calibration, we have obtained long slit spectra for a number 
of radial velocity standard stars.
In addition, we have taken integrated spectra of several Galactic Globular 
Clusters (GGCs) in order to calibrate metallicities. 
The spatial extent of these globular clusters on the sky 
(mean core radii $\sim$ 40 arcsec) required us to synthesize an aperture in 
order to obtain a representative integrated spectrum. 
We have therefore scanned the cores of the globular clusters over a typical 
range of 90 arcsec. Combined with the WHT observations, this yields
high S/N spectra ($\sim$ 500 at 5000 \AA) of five GGCs in 
the metallicity range -2.24 $<$ [Fe/H] $<$ -0.29 dex with one overlap, namely 
NGC~6356.
We summarize our calibration objects in Table~\ref{tab:standards}.

\begin{table}
\begin{center}
\caption{\small{Programme standard stars and GGCs for radial velocity and 
metallicity calibration.}}
\label{tab:standards}
\begin{tabular}{llll}

\hline 

   ID & Object & Spectral Type & [Fe/H]$^{1}_{\rm{Z}}$\\ 

\hline

NGC~6341 & GGC & F2 & -2.24 \\
NGC~7078 & GGC & F3-4 & -2.15 \\ 
NGC~6205 & GGC & F6 & -1.65 \\
NGC~6356 & GGC & G3 & -0.62 \\
NGC~6553 & GGC & G4 & -0.29 \\
HD~90861 & Star & K2III & - \\
HD~102494 & Star & G9IV & - \\
HD~112299 & Star & F8V & - \\
HD~132737 & Star & K0III & - \\
HD~171232 & Star & G8III & - \\
HD~208906 & Star & F8V-VI & - \\

\hline

\end{tabular}
\end{center}
$^{1}$ Abundances from Zinn (1985), henceforth denoted [Fe/H]$_{\rm {Z}}$.  

\end{table}

The majority of the data reduction was performed using the package 
\textsc{doslit} in \textsc{iraf} together with other standard tasks.
Object frames were trimmed and bias-subtracted, and any bad pixels were
cleaned by interpolating across adjacent columns. 
A second-order polynomial fit to these data produced residuals of order 
$\sim$ 0.1 \AA\ in the wavelength calibration.
The spectra were then optimally extracted and sky-subtracted using  
a linear least-squares fit to the background sky. 
Finally, the spectra were rebinned with a step size of 2 \AA\ on to a 
logarithmic wavelength scale over the range 3800 -- 5500 \AA. 
The resultant spectra typically possessed S/N ratios of 4 -- 10.

\section{Analysis}

\subsection{Radial Velocities}

The Fourier cross-correlation task \textsc{fxcor} was used for determining 
radial velocities from the spectra. 
The six velocity templates were de-redshifted by their literature 
values and then individually cross-correlated against each of the 
candidate cluster spectra. 
An 'r' value of 2.5 (Tonry \& Davis 1979) was set as the lower threshold
for reliable measurement; below this any velocities returned, however 
plausible, were removed from further analysis. 
The final velocities of the globular clusters were taken to be the mean velocity
weighted by the cross-correlation peak height of each template. 
A heliocentric correction was then applied to these velocities.
In addition to the formal errors returned by the cross-correlation task, 
an estimate of the uncertainties may be made from the overlap between common 
objects in different masks. 
Between the data from Sharples \etal (1998) and the CFHT velocities 
there are a total of 13 overlaps. 
The mean difference of the sample is -17 kms$^{-1}$ with an inferred velocity
uncertainty for a single measurement of 78 kms$^{-1}$.

We take a velocity range 300 $\le V_{\rm{h}} \le$ 2000 kms$^{-1}$ as being
representative of globular clusters associated with NGC~4472, which
has a heliocentric velocity of 961 kms$^{-1}$ (Sandage \& Tammann 1981).
This is consistent with the mean velocity of our globular cluster 
sample of 990  $\pm$ 26 kms$^{-1}$, 
which has a velocity dispersion of 314 kms$^{-1}$. See Zepf \etal (in 
preparation) for further analysis of the globular cluster kinematics.

\subsection{Co-adding the Spectra}

Whilst the quality of our individual spectra is adequate to obtain 
radial velocities, it is insufficient for reliable
line-strength analysis. For any believable measurement of equivalent widths
of absorption lines, a method of co-addition of the spectra was required 
so as to improve the S/N ratio.

\begin{figure}
\epsfysize 3.0truein
\hfil{\epsffile{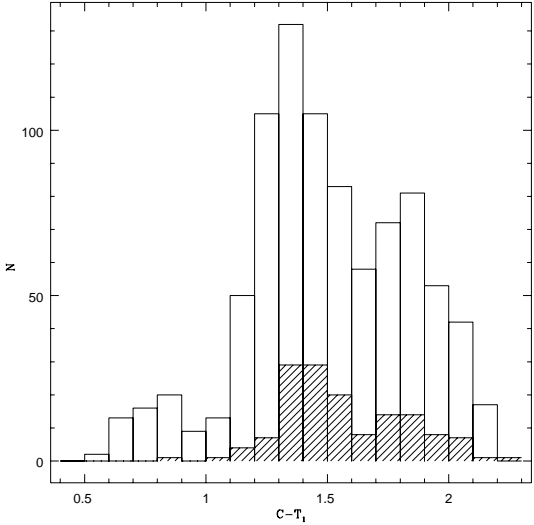}}\hfil
\caption{\small{Colour distribution for globular cluster candidates with 
19.5 $\leq V \leq$ 22.5 and 0.5 $< C-T_{1} <$ 2.2 from 
Geisler \etal (1996). 
The open histogram is for the full sample of 860 candidates; the shaded 
histogram is for the 131 globular clusters for which radial velocities have been  
obtained.}} 
\label{fig:bimodal}
\end{figure}

Since the broadband colours of the globular clusters primarily reflect
their metallicities (on the assumption that these are old stellar 
populations, $\tau \ga$ 8 Gyr), blue globular clusters should be metal-poor, 
and become progressively more metal-rich as they redden. 
The cluster spectra were therefore assigned bins on the basis of 
their $C-T_{1}$ colours from the photometry of Geisler \etal (1996).
The globular cluster catalogue of Geisler \etal (1996) is bimodal in colour, 
and the subset of brightest globular clusters selected from this catalogue 
were 
chosen so as to reflect this bimodality, albeit with reduced cluster numbers.
Fig.~\ref{fig:bimodal} shows the sample of 860 cluster candidates from
 Geisler \etal (1996) (open histogram) and those for which radial 
velocities have been obtained and subsequently used in the metallicity 
analysis (shaded histogram).
Although we now have velocities for 141 globular clusters, 10 of these 
were from the original sample of Mould \etal (1990) and these spectra were 
unavailable for further analysis.

Due to this bimodality in these data, creating bins of a fixed width in 
colour leads to significantly different numbers of spectra in each bin.
Spectra resulting from such binning are not readily comparable since their 
signal-to-noise (S/N) varies as a function of the number of spectra per bin.
The bins were therefore constructed by
assigning approximately equal numbers of spectra to each bin, yielding final
spectra of similar S/N, but with different widths in colour. 
In order to determine the number of spectra per bin acceptable for  
reliably measuring line-strengths, the following procedure was adopted.
The Spectral Energy Distributions (SEDs) of Worthey (1994) (see $\S5.2$)
at a fixed age and range of metallicities were degraded to the typical 
S/N of the NGC~4472 globular cluster spectra. 
This was achieved by adding random noise to the spectra 
until they possessed a distribution in S/N equivalent to that 
of the NGC~4472 globular cluster data.
Subsequent spectra were then co-added and their line-strength indices 
measured (see $\S4.3$ for discussion of index measurements).
This process was reiterated until a sufficient number of realisations 
produced stable measurements in all indices, that were consistent 
with the indices measured for the original, undegraded spectrum.

\begin{figure}
\epsfysize 3.0truein
\hfil{\epsffile{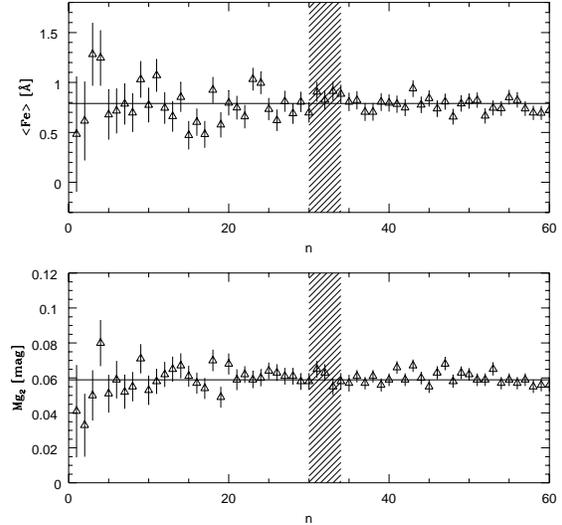}}\hfil
\caption{\small{Line-strength measurements of $<$Fe$>$ (top panel) 
and Mg$_{2}$ of the degraded and co-added SEDs of Worthey (1994).
Each set of realisations is represented by an open triangle, with its 
associated statistical error. The shaded region indicates the location
where the line-strength measurements were found to be stable.
The solid horizontal line in each panel indicates the value of the index 
measurement for the original SED. Note that $<$Fe$>$ is measured in units
of Angstroms, whereas Mg$_{2}$ is expressed as a magnitude.
}}
\label{fig:dispersion}
\end{figure}

An example of the results of this exercise is shown in 
Fig.~\ref{fig:dispersion} for the $<$Fe$>$\footnote{$<$Fe$>$ is defined as
the mean of the Fe5270 and Fe5335 indices in Gonzal$\acute{e}$z (1993).} 
and Mg$_{2}$ indices. 
Each filled circle represents a set of \emph{n} realisations, with their 
associated statistical errors, derived from the S/N of each 
spectrum obtained by summing over pixels along the slit within a fixed 
wavelength range ($\sim$ 4545 -- 5500 \AA).
The shaded column denotes where the number of co-added spectra were found 
to be sufficient to produce a stable index measurement.
In this case, the degraded SED is that predicted
by the Worthey (1994) models for a 17 Gyr stellar population with
[Fe/H] = -1.7 dex. The indices measured from the co-added spectra (at \emph{n}
 = 32) in Fig.~\ref{fig:dispersion} are 0.065 $\pm$ 0.011 mag for Mg$_{2}$, 
and 0.85 $\pm$ 0.30 \AA\ for $<$Fe$>$, in excellent agreement with the 
models which predict 0.06 mag for and 0.79 \AA\ for $<$Fe$>$.

This procedure, repeated for different index features, showed no 
evidence of systematic effects. 
Initially, four colour bins where selected (set by the number of 
spectra required per bin), the first (bluest) consisting
 of 35 spectra, the remaining three each comprising of 32 spectra.
Prior to co-addition, the spectra were transferred to the rest-frame 
using their measured radial velocities.
This procedure is essential for any measurement of line indices using 
fixed-wavelength bandpasses (e.g. Brodie \& Huchra 1990). 

\subsection {Line Indices}

Numerous systems for measuring line-strengths in stellar populations  exist, 
all with their respective merits and failings. However, the key features 
considered important here are the size and availability of the observational 
database, and the range of stellar models with which to calibrate these data. 
In this respect, the Lick system
(Burstein \etal 1984; Faber \etal 1985; Gorgas \etal 1993)
is our system of choice.

The absorption-line indices were measured following
a prescription similar to that described in Brodie \& Huchra (1990), 
which is based upon the Lick/IDS system of bandpasses as presented in 
Burstein \etal (1984). Each index is fully described by two pseudocontinuum 
regions and one feature bandpass region. 

For atomic lines, the equivalent width is defined as:

\begin{equation}
\label{eq:ew}
EW =\int^{\lambda_{2}}_{\lambda_{1}}\left(1-\frac{F_{I\lambda}}{F_{C\lambda}}\right)~d\lambda
\end{equation}       

where $\lambda_{1}$,$\lambda_{2}$ are the wavelength limits of the passband, 
and $F_{I\lambda}$,$F_{C\lambda}$ are the flux per unit wavelength of the
measured feature and the continuum respectively.

Molecular line-strengths are similarly defined, but are measured in magnitudes:

\begin{equation}
\label{eq:mag}
I =-2.5~\rm{log}\left[\left(\frac{1}{\lambda_{2}-\lambda_{1}}\right)~\int^{\lambda_{2}}_{\lambda_{1}} \frac{F_{I\lambda}}{F_{C\lambda}}~d\lambda\right]
\end{equation}

\begin{figure}
\epsfysize 3.0truein
\hfil{\epsffile{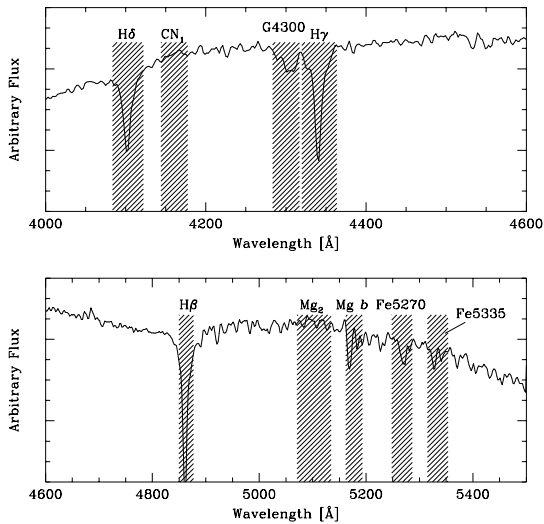}}\hfil
\caption{\small{Location of the feature bandpasses measured in our data.
 Shaded regions show the width and position of the 
features, which are over-plotted on to the normalised spectrum of the 
metal-poor globular cluster M15.}}
\label{fig:indices}
\end{figure}

We have also measured two more Balmer indices, H$\gamma_{\rm{A}}$ and
H$\delta_{\rm{A}}$, defined by Worthey \& Ottaviani (1997). The subscript
 'A' denotes the wider of their four indices (they also define
H$\gamma_{\rm{F}}$ and H$\delta_{\rm{F}}$ which are $\sim$ 20 \AA\ wide) which
were principly designed to measure the absorption from A-stars.

For comparison with the stellar population models of 
Worthey (1994), we use the index definitions given in Worthey \etal (1994) 
and Worthey \& Ottaviani (1997).
To compare with the empirical calibrations of Brodie \& Huchra (1990), we
also measure indices using the index definitions given in Brodie \& Huchra (1990), which differ largely in their placing of the continuum. 
Fig.~\ref{fig:indices} illustrates the locations of the feature 
bandpasses measured in this study, the regions of pseudocontinua are 
omitted for clarity.
 
The Lick/IDS system is based upon a stellar library which is \emph{not} 
flux-calibrated, and its resolution (FWHM) varies as a function of wavelength. 
Our data have a fixed spectral resolution of $\sim$ 5.5 \AA\ (FWHM) and 
have also not been flux-calibrated (due to the difficulties of fluxing multi-slit data).
Consequently, it is difficult to accurately transform our measurements on to 
the Lick System and care must be taken, particularly with regard to
the continuum slope (e.g. Worthey \& Ottaviani 1997; Cohen \etal 1998).

\begin{table}
\begin{center}
\caption{\small{Dependence of spectral resolution on wavelength for the 
Lick/IDS.}}
\label{tab:wo}
\begin{tabular}{cc}

\hline 

Wavelength & Resolution (FWHM)\\
$[$\AA$]$ & [\AA] \\

\hline

4000   ....... & 11.5 \\
4400   ....... & 9.2 \\
4900   ....... & 8.4 \\
5400   ....... & 8.4 \\
6000   ....... & 9.8 \\

\hline

\end{tabular}
\end{center}
\end{table} 

The first step in approximating to the Lick system is to convert the 
resolution of our data to that of the IDS. In Table~\ref{tab:wo} we 
reproduce the data given in table 8 of Worthey \& Ottaviani (1997), 
illustrating the dependence of spectral resolution upon wavelength for the 
IDS.
For our multislit data, the useful spectral range is 3800 \AA\ -- 
5500 \AA, covering a Lick resolution from approximately 11.5 \AA\ 
to 8.4 \AA.
We first divide our co-added spectra into 200 \AA\ intervals, and the 
resolution required to best match the Lick/IDS spectra within that 
interval was taken from Table~\ref{tab:wo}.
Our spectra were then degraded from a resolution of 
$\delta\lambda_{\rm{N4472}}$ to the IDS at $\delta\lambda_{\rm{IDS}}$ 
by convolution with a Gaussian kernel of width $\sigma$, given according 
to Eqn.~(\ref{eq:gauss}).

\begin{equation}
\label{eq:gauss}
\sigma =\frac {\sqrt{(\delta\lambda_{\rm{IDS}}^{2}-\delta\lambda_
{\rm{N4472}}^{2})}}{2.35 * \rm{\AA}~\rm{pix}^{-1}} 
\end{equation}       

The indices of the unbroadened spectra were then compared to those of the
Lick resolution spectra to check for any possible systematic differences.
The broader indices were largely unaffected, for example Mg$_{2}$,
 which has a width of 42.5 \AA, was unchanged. 
However Mg $b$, a narrower index (width 32.5 \AA) 
was systematically lower by approximately 0.14 \AA\ when measured 
from the broadened spectra. 

The second consideration is the effect of the instrumental response 
curve upon the continuum slope of the indices. 
We have simulated the effect of varying the local spectral
slope of each feature bandpass, by altering the 
\emph{global} continuum shape of the spectra.
A polynomial was fitted to each of our co-added spectra, representing 
the full range of continuum shape in our data.
This entire set of artificial 'response functions' were then added
to a normalised standard spectrum and the indices of each 
were measured and the results compared.
Table~\ref{tab:responses} shows the effect upon the index measurements 
resulting from this variation in the continuum. 

\begin{table}
\begin{center}
\caption{\small{Effect of the continuum shape on index measurements of 
features in HD 102494. For comparison purposes, we have converted the
molecular indices CN$_{1}$ and Mg$_{2}$ into equivalent widths of
absorption.}}
\label{tab:responses}
\begin{tabular}{lcccccccc}

\hline 

Feature & Mean & Median & $\sigma$ & Width \\

 & [\AA] & [\AA] & [\AA] & [\AA]\\

\hline

CN$_{1}$  & 0.193 & 0.289 & 0.42 & 35 \\
G4300  &  3.017 & 2.986 & 0.10 & 35 \\ 
H$\beta$  & 0.808 & 0.801 & 0.02 & 28.75\\
Mg$_{1}$  &  0.358 & 0.297 & 0.54 & 65 \\
Mg$_{2}$  &  2.396 & 2.322 & 0.20 & 42.5 \\
Mg $b$  &  1.542 & 1.545 & 0.02 & 32.5 \\
Fe5270 & 1.335 & 1.399 & 0.02 & 40 \\
Fe5335 & 0.982 & 0.984 & 0.01 & 40 \\
H$\delta_{A}$ & -1.739 & -1.694 & 0.19 & 38.75 \\
H$\gamma_{A}$ & -3.016 & -3.041 & 0.10 & 43.75 \\
\hline

\end{tabular}
\end{center}
\end{table} 

As Table~\ref{tab:responses} indicates, the majority of the 
features remain relatively stable. However, it is apparent that 
the CN$_{1}$ and Mg$_{1}$ features are very sensitive to small shifts in 
the slope of the local continuum and cannot be trusted in these current data.
To a lesser extent, Mg$_{2}$ and H$\delta_{\rm{A}}$ are also susceptible to
changes in the continuum level. 
In view of this we omit the CN$_{1}$ and Mg$_{1}$ from further
analysis, and recognize the behaviour Mg$_{2}$ and H$\delta_{\rm{A}}$.   
We note that \emph{extreme care must be taken when measuring indices under 
the Lick system if the continuum has not been corrected to that of the IDS. 
This is especially true for the broader features.}

Errors in the velocities of the individual spectra had no significant 
effect upon the line indices measured for our co-added spectra. 
Our uncertainties in the radial velocities velocities are $\le$ 
100 kms$^{-1}$, 
translating to approximately half a pixel at the resolution of our data.
Similarly, we applied no correction for line broadening due to any 
\emph{internal} velocity dispersion of the globular clusters. 
Internal stellar velocity dispersions of globular clusters are small, 
typically $\sim$ 10 kms$^{-1}$ or less (e.g. Meylan \& Heggie 1997).

\subsection{Uncertainties in the Indices}

\begin{figure}
\epsfysize 3.0truein
\hfil{\epsffile{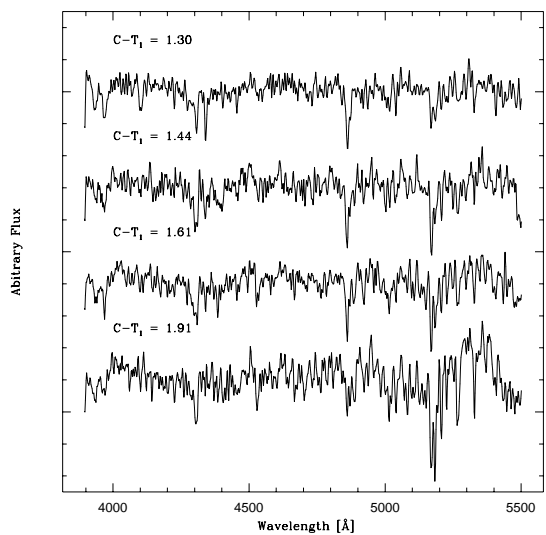}}\hfil
\caption{\small{Representative co-added spectra of the globular clusters 
obtained from the bootstrap procedure. 
The spectra become progressively redder from top to bottom. 
The mean colours in $C-T_{1}$ are 1.30, 1.44, 1.61 and 1.91 respectively. 
The spectra have been divided through by a fifth-order polynomial, and 
have been shifted in the \emph{y}-axis for clarity.
}}
\label{fig:binned_specs}
\end{figure}

The value of any line-strength measurement is dependent upon a 
suitable calculation of the associated errors. 
For an old stellar population at roughly solar metallicity, an error 
as small as 0.2 \AA\ in the Mg $b$ index translates into an 
uncertainty of $\sim$ 0.2 dex in metallicity, and an uncertainty of 
0.2 \AA\ in the H$\beta$ index corresponds to an error of $\sim$ 3 Gyr in age.
The main sources of uncertainty in the index measurements of the co-added 
spectra stem from systematics in the binning process and any 
intrinsic spread in the chemical composition of the globular clusters. 
The latter is inherent in these data; the spectra comprise of colour bins  
of a finite width which can only be adjusted through alterations in this 
binning. 
These uncertainties may be best estimated through the use of a bootstrap 
technique, which takes into account the stochastic nature of co-adding 
individual, low S/N spectra. The method we use is quite straightforward:  
each of our colour bins contain \emph{n} spectra, and each spectrum is 
assigned a number from 1 to \emph{n}. 
A spectrum is then selected randomly from the bin by its corresponding 
number and placed into a new pool which holds the spectra to be combined. 
This process is then repeated until the resulting in a pool contains 
\emph{n} randomly selected spectra from the original bin.
The spectra are selected \emph{with replacement}, so the bin is never 
depleted and the new pool is likely to contain several repetitions of 
the same spectrum. 
The spectra contained within the pool are then combined to produce a 
final composite spectrum, which then has its line indices measured.
 This description comprises one realisation and is repeated
many times so as to produce many index measurements of the same feature.
Typically 150 realisations were performed for each colour bin. 
We take the final index value to be the mean of the distribution of
measurements for each bin and its associated uncertainty as 
the standard deviation of this distribution.
In addition, we have also measured the flux-weighted colours for
each bin. This is preferable over a magnitude weighting
scheme, since the amount of light received through each slit on to the 
detector is not only a function of an individual globular clusters' 
brightness, but also of its position in the slit. 
Fig.~\ref{fig:binned_specs} shows representative co-added spectra for the 
four colour bins, which have the mean flux-weighted colours of
$C-T_{1}$= 1.30, 1.44, 1.61 and 1.91 respectively.
Metal lines clearly become progressively stronger with redder colours
(e.g. Mg $b$ feature at $\sim$ 5180 \AA), whilst the Balmer lines weaken 
(e.g. H$\beta$ at 4860 \AA\ and H$\gamma_{\rm{A}}$ at 4340 \AA).

\begin{figure}
\epsfysize 3.0truein
\hfil{\epsffile{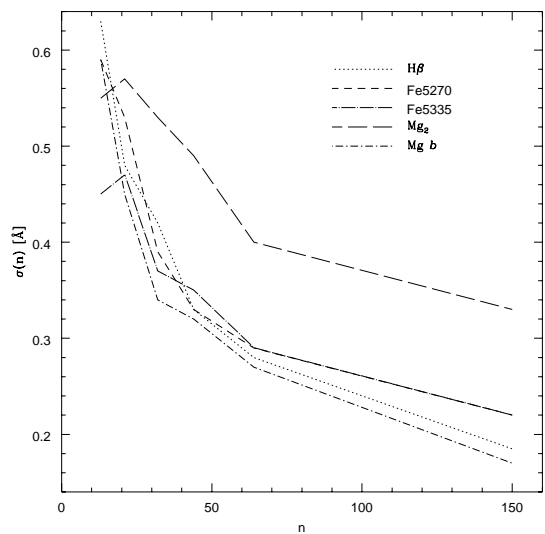}}\hfil
\caption{\small{The dependence of the variance in the measured  
indices of the co-added spectra, upon the number of spectra per bin.
}}
\label{fig:binsize}
\end{figure}

We have extensively tested the effects of altering the bin size
for our globular cluster data. We have varied \emph{n}, the number 
of spectra per bin, and repeated the above procedure determining the
variance in the indices in each case. Fig.~\ref{fig:binsize} shows
the dependence of $\sigma$ upon \emph{n} for our co-added spectra.
The uncertainties in all of the indices behave in a similar fashion,
with a steep initial gain in accuracy of $\sim$ 0.02 \AA\ / spectrum, 
and then flattening considerably with increasing 
\emph{n}. The uncertainties in the Mg$_{2}$ index are $\sim$ 0.1 \AA\
greater than for the other indices, which we largely largely
attribute to its sensitivity to variations in the continuum slope. 
Fig.~\ref{fig:binsize} indicates that we may expect an uncertainty
in most of the indices of 0.3 \AA\ -- 0.4 \AA\ and 0.5 \AA\ 
(0.013 mag) in Mg$_{2}$  for \emph{n} $\simeq$ 30.

\subsection{Results}

The indices measured for the four cluster bins using the bandpass 
definitions in Worthey \etal (1994) and Worthey \& Ottaviani (1997)
are given in Table~\ref{tab:results}. 
Associated uncertainties obtained from the bootstrap are tabulated in 
alternate rows. 
Errors for the mean bin colour are the standard deviation of the
distribution of the flux weighted $C-T_{1}$ colours from the bootstrap 
procedure.
Column 3 of Table~\ref{tab:results} gives the metallicities predicted by
the $C-T_{1}$ colours using the linear relation of Geisler \& Forte
 (1990):

\begin{table*}
\begin{center}
\caption{\small{Measured indices for our four colour bins of NGC~4472 
globular clusters. Column 2 gives the flux weighted mean $C-T_{1}$ colour for 
the bin. The metallicity given in column 3 is derived from these colours 
using Eqn.~(\ref{eq:doug}), the relation of Geisler \& Forte (1990) and is
henceforth denoted [Fe/H]$_{\rm{GF}}$. 
Uncertainties derived from the bootstrap are given in alternate rows. }}
\label{tab:results}
\begin{tabular}{lccccccccccc}
\hline 

Bin & C-T$_{1}$ & [Fe/H]$_{\rm{GF}}$ & G4300 & H$\beta$ & Mg$_{2}$ & Mg $b$ & Fe5270 & Fe5335 & H$\delta_{\rm{A}}$ & H$\gamma_{\rm{A}}$ \\
 & [mag] & (dex) & [\AA] & [\AA] & [mag] & [\AA] & [\AA] & [\AA] & [\AA] & [\AA]\\
\hline

1 & 1.30 & -1.34 & 2.030 & 2.452 & 0.072 & 1.648 & 1.573 & 1.093 & 2.497 & 0.374\\
... & $\pm$ 0.09 & 0.21 & 0.397 & 0.330 & 0.011 & 0.303 & 0.275 & 0.274 & 0.815 & 0.566\\
2 & 1.44  & -1.00 & 3.590 & 1.956 & 0.114 & 2.287 & 1.778 & 0.942 & -0.466 & -1.607\\
... & $\pm$ 0.07 & 0.16 & 0.550 & 0.430 & 0.011 & 0.295 & 0.436 & 0.317 & 1.346 & 0.582\\
3 & 1.61 & -0.61 & 4.017 & 1.630 & 0.126 & 2.115 & 1.706 & 1.432 & 0.018 & -2.763\\
... & $\pm$ 0.06 & 0.14 & 0.469 & 0.244 & 0.008 & 0.206 & 0.289 & 0.179 & 0.896 & 0.753\\
4 & 1.91 & 0.10 & 6.417 & 1.225 & 0.240 & 4.347 & 3.195 & 2.770 & -2.052 & -6.354\\
... & $\pm$ 0.11 & 0.26 & 0.943 & 0.410 & 0.019 & 0.359 & 0.331 & 0.468 & 1.913 & 1.658\\

\hline

\end{tabular}
\end{center}
\end{table*} 

\begin{equation}
\label{eq:doug}
{\rm [Fe/H]} = 2.35 * (C-T_{1}) - 4.39
\end{equation} 

Eqn.~(\ref{eq:doug}) is based upon a calibration of 48 GGCs using the 
photometry of Harris \& Canterna (1977) and [Fe/H] values from Zinn (1985) 
and Armandroff \& Zinn (1988). The range of metallicity spanned by 
this calibration is -2.5 $<$ [Fe/H] $<$ -0.25 dex.

\section{Globular Cluster Metallicities and Ages}

\subsection{Fiducial Globular Clusters}

As a quantitative comparison, and in order to calibrate the stellar 
population models that we will use for obtaining abundances and ages, 
we have obtained trailed, long-slit integrated spectra for five GGCs.
We supplement this with data from Brodie \& Huchra (1990) and
Cohen \etal (1998). 
The measured Lick indices for our combined WHT and CFHT 
sample of GGCs are given in Table~\ref{tab:galactic}. Zinn (1985) quotes
a typical uncertainty in the metallicities of the GGCs (column 11 in 
Table~\ref{tab:galactic}) of 0.15 dex.

\begin{table*}
\begin{center}
\caption{\small{Measured Lick/IDS indices for GGCs 
from the WHT and CFHT observing runs. Uncertainties are derived from the
quadrature sum of the Poisson statistics and the rms between overlapping
measurements.}}
\label{tab:galactic}
\begin{tabular}{llccccccccc}
\hline 
ID & Run & G4300 & H$\beta$ & Mg$_{2}$ & Mg $b$ & Fe5270 & Fe5335 & H$\delta_{\rm{A}}$ & H$\gamma_{\rm{A}}$ & [Fe/H]$_{\rm{Z}}$ \\
 & & [\AA] & [\AA] & [mag] & [\AA] & [\AA] & [\AA] & [\AA] & [\AA] & (dex) \\
\hline
NGC~6341  & WHT  &  0.83  &  2.63  &  0.013  &  0.56  &  0.47  &  0.34  &  3.94  &  2.37  & -2.24\\
... & ... & $\pm$  0.72  &  0.05  &  0.009  &  0.07 &  0.15 & 0.03 & 0.13 & 0.19 & ...\\  
NGC~7078  & CFHT &  0.37  &  3.10  &  0.018  &  0.53  &  0.96  &  0.37  &  4.71  &  3.43  & -2.15\\
... & ... & $\pm$  0.06 &  0.03 & 0.002 &  0.04 & 0.04 &  0.04 &  0.07 &  0.06 & ... \\
NGC~6205  & WHT  &  2.14  &  2.31  &  0.045  &  1.00  &  0.92  &  0.78  &  3.17  &  0.71  & -1.65\\
... & ... & $\pm$ 0.82 &  0.08  &  0.003 &  0.12 &  0.02   &  0.05 &  0.13 &  0.13 & ...\\ 
NGC~6356  & CFHT &  4.98  &  1.73  &  0.151  &  3.13  &  1.80  &  1.61  &  -0.52  &  -3.79  & -0.62\\
... & ... & $\pm$  0.38 &  0.19 &  0.011 & 0.09 & 0.11 &  0.10 & 0.16 & 0.15 & ... \\
NGC~6356  & WHT  &  4.72  &  1.57  &  0.174  &  3.16  &  1.94  &  1.49  &  0.08  &  -3.80  & -0.62\\
... & ... & $\pm$  0.20 &  0.17  &  0.012 &  0.08 &  0.11 & 0.12 & 0.10 & 0.09 & ...\\  
NGC~6553  & CFHT &  5.08  &  1.55  &  0.244  &  4.11  &  3.13  &  2.38 &  -0.90 & -5.59 & -0.29\\
... & ... & $\pm$ 0.17 & 0.30 & 0.002 & 0.15 & 0.04 & 0.07 &  0.18 & 0.21 & ... \\
\hline

\end{tabular}
\end{center}
\end{table*}

\begin{figure}
\epsfysize 3.0truein
\hfil{\epsffile{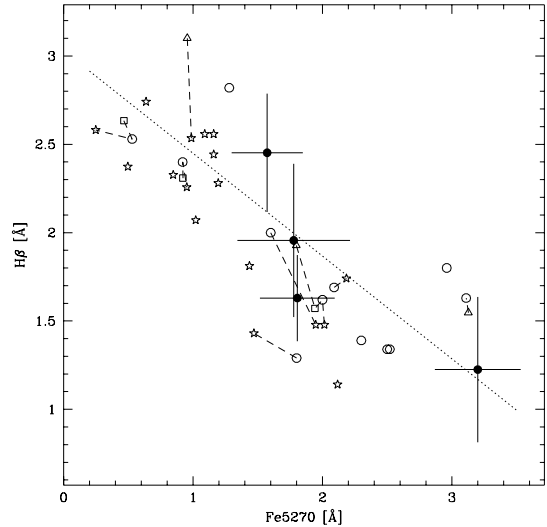}}\hfil
\caption{\small{H$\beta$ index as a function of Fe5270. Solid circles
with error bars are for the co-added NGC~4472 cluster data. 
Open symbols are GGCs: CFHT - triangles, 
WHT - squares, circles indicate the data of Cohen \etal (1998) and stars 
represent data from Brodie \& Huchra (1990). Dashed lines join 
globular clusters common between the datasets. 
The dotted line indicates the formal linear fit to our co-added NGC~4472
 data.}}
\label{fig:fe52_hbeta_qual}
\end{figure}

\begin{figure}
\epsfysize 3.0truein
\hfil{\epsffile{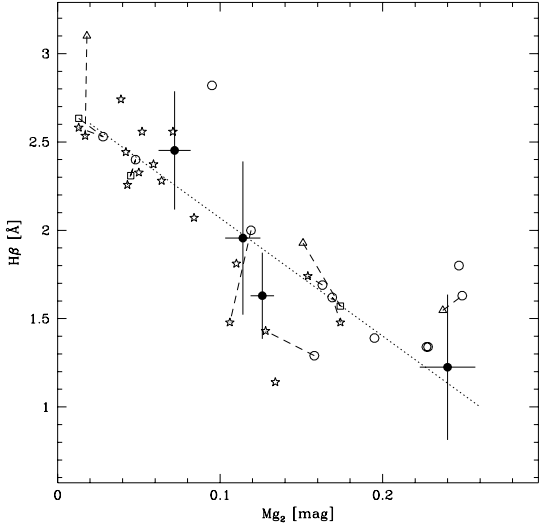}}\hfil
\caption{\small{The H$\beta$ index against Mg$_{2}$. 
Symbols are the same as for Fig~\ref{fig:fe52_hbeta_qual}.}}
\label{fig:mg2_hbeta_qual}
\end{figure}

\begin{figure}
\epsfysize 3.0truein
\hfil{\epsffile{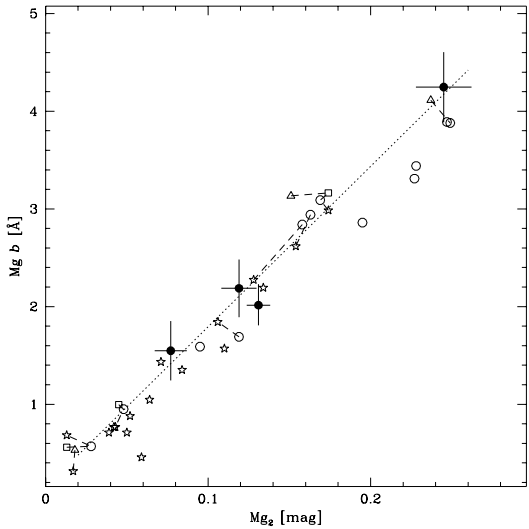}}\hfil
\caption{\small{The Mg $b$ index as a function of Mg$_{2}$. 
 Symbols are the same as for Fig~\ref{fig:fe52_hbeta_qual}.}}
\label{fig:mg2_mgb_qual}
\end{figure}

\begin{figure}
\epsfysize 3.0truein
\hfil{\epsffile{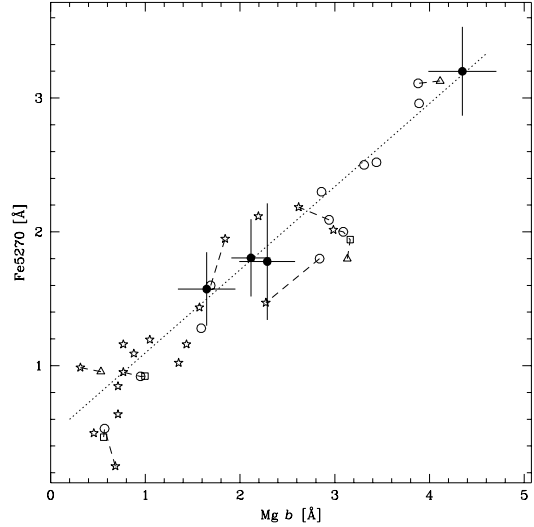}}\hfil
\caption{\small{Fe5270 index versus Mg $b$. 
Symbols are the same as for Fig~\ref{fig:fe52_hbeta_qual}.}}
\label{fig:mgb_fe52_qual}
\end{figure}

In Figs.~(\ref{fig:fe52_hbeta_qual} -- \ref{fig:mgb_fe52_qual}) we 
plot our co-added NGC~4472 data together with the combined sample of GGCs 
for which we have line-strength measurements. The linear fits in the
figures applied to our data are not physically motivated, but merely serve 
as a comparison to the GGCs.
Inspection of these figures reveals several points.
As expected, the H$\beta$ index increases with the 
decreasing strength of the more metallicity-sensitive lines. 
The co-added cluster data of NGC~4472 falls within the
metallicity and age ranges of the GGCs. 
The metal-rich ends of these data coincide at approximately solar, 
but the GGCs extend out to much lower metallicities. 
The mean line of the NGC~4472 globular clusters
traces that of the GGC distribution fairly closely.
The index-index plots in Fig.~\ref{fig:mg2_mgb_qual} and 
Fig.~\ref{fig:mgb_fe52_qual} are well-correlated with each other, 
with no obvious evidence of systematic offsets between the different indices.
Although there is only one overlap between the two observing runs, 
namely NGC~6356, the agreement is good in this case.
Five of our GGCs are in common with the combined dataset of GGCs from
Brodie \& Huchra (1990) and Cohen \etal (1998), and in most cases 
these data are in good agreement. However, our measurement of H$\beta$ 
for NGC~7078 (M92) is $\sim$ +0.5 \AA\ offset from the value
measured by Cohen \etal (1998), and the reason for this is
unclear.

In summary: the reddest bin of our globular clusters has a similar 
metallicity to the most metal-rich GGCs. 
The bluest bin is significantly more metal-rich than the bluest 
GGCs. The indices act together in the expected
sense, metal-lines strengthen towards the red and the Balmer 
indices weaken. 
The co-added data are self-consistent within the bootstrapped 
uncertainties, which gives us confidence in our data and
error estimates.

\subsection{SSP models}

\begin{figure}
\epsfysize 3.0truein
\hfil{\epsffile{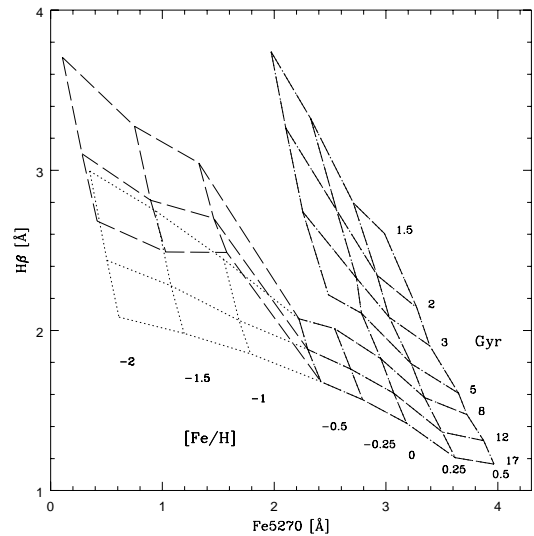}}\hfil
\caption{\small{Effect of horizontal branch morphology on the 
Fe5270--H$\beta$ grid of the SSP models of Worthey (1994). 
The horizontal scale indicates the positions of constant lines
of metallicity (running almost vertically here), ranging from 1/100 to 3 
times solar. 
The approximately horizontal grid lines represent isochrones ranging from 1.5 
to 17 Gyr (read from top to bottom).
Faint dotted lines indicate the predictions of the original models 
assuming a red clump, long dashed lines 
show the result of including extended HBs following the observed 
behaviour of GGCs.}}
\label{fig:grids}
\end{figure}

To derive the ages and metallicities of the NGC~4472 globular 
clusters, we compare our data with single age, single metallicity 
stellar population models (SSPs).
These model aggregates of coeval stars
with homogeneous metal content, i.e. an idealised globular cluster.
The SSP models considered here are the on-line versions of  
Worthey (1994) which can come in several variants, and may be
 fine--tuned from the web-page of Dr. 
Worthey.\footnote{http://astro.sau.edu/$\sim$worthey/}
For our purposes, we restrict ourselves to models with a Salpeter initial
mass function, with the assumption of a single, instantaneous burst of 
star formation. 
However, the grids used in this study differ from the original 
Worthey (1994) models in one important respect, which is in their 
treatment of the horizontal branches (HBs) for metal-poor globular clusters. 
The morphology of the HB now follows the observed
behaviour of the GGCs, in that they become more 
extended toward the blue for more metal-poor globular clusters, as opposed to 
assuming a red clump at the base of the HB irrespective of metallicity. 
The GGCs M3 and M92 were used 
as templates for this behaviour (e.g. Aaronson \etal 1978 - for further 
discussion see Worthey 1993). 
This change significantly increases the predicted 
H$\beta$ index by upwards of 0.5~\AA\ at [Fe/H] $\leq$ -0.5 dex. 
Fig.~\ref{fig:grids} illustrates this effect
and its importance with regard to the predicted ages for 
globular clusters, of which a substantial population have sub-solar 
abundances. 
A globular cluster with a measured Fe5270 index of 1.5 \AA\ 
and H$\beta$ index of 2.4 \AA\ would be assigned an age of 8 Gyr according to 
the old models, but is predicted to be $\geq$ 17 Gyr old by the models 
using the new HB morphology.
 
As a case in point, it is interesting to compare the H$\beta$
indices of GGCs at the same metallicity, but with differing
HB morphologies. 
One of our calibrating GGCs is M13 (NGC~6205), for which we have measured 
H$\beta$ = 2.40 \AA, and has a HB ratio of approximately unity 
(effectively it has the same number of stars both blueward and 
redward of the RR Lyrae gap). 
Amongst the sample of  Cohen \etal (1998) is the GGC M3 (NGC~5272), 
for which they obtain H$\beta$ = 2.31 \AA, and has an extremely
blue HB. 
Since both of these GGCs have Fe/H $\sim$ -1.65, this would imply that HB 
morphology has little or no effect upon H$\beta$, contrary to the predictions
of the above SSP models.
Clearly, a high quality, homogeneous data set is required to investigate 
whether the horizontal branch may affect the Balmer indices of globular 
clusters at the same metallicity. 

The model grids cover a metallicity-age parameter space of 
-0.5 $\leq$ [Fe/H] $\leq$ 0.5 with isochrones of 1.5  $\leq \tau \leq$ 
17 Gyr.
To cover old, metal-poor populations, they have been subsequently extended to
bracket -2.0 $\leq$ [Fe/H] $\leq$ 0.5 for ages 8  $\leq \tau \leq$ 17 Gyr.
The orthogonality of the index-index plots indicates the degree of
decoupling between age and metallicity in the models.

The assumption of a red clump in the Worthey (1994) models 
possibly explains the findings of Cohen 
\etal (1998) that formally their metal-rich GGCs
are \emph{older} than the metal-poor ones, the opposite of 
what one might expect. This new treatment of the HBs in the models 
predicts similar ages for the metal-rich and metal-poor GGCs.

\subsection{Model Predictions}

\subsubsection{Metallicities of the Globular Clusters}

We calibrate the SSP models using the indices measured for the combined
sample of GGCs.
The metallicities predicted by the models of Worthey (1994) 
must first be placed on to the metallicity scale of our 
GGCs.
We use the most widely used metallicity scale for GGCs, 
which is that of Zinn (1985). This is primarily 
based on a number of narrow-band photometric indices
sensitive to the ultraviolet blanketing in the integrated
light of the globular cluster spectral energy distribution. Whilst there is 
increasing evidence that this metallicity scale is non-linear
(e.g. Carretta \& Gratton 1997), we adopt it since it remains 
self-consistent and the empirical calibrations of Brodie \& Huchra (1990) 
are based upon this (see below).

We first obtain [Fe/H] on the Worthey (1994) scale as a function of
the Fe5270, Mg $b$ and Mg$_{2}$ indices for our
entire sample of GGCs. The newly assigned metallicities for the globular 
clusters are then 
compared to those of Zinn (1985) in order to derive a transformation 
function from the model metallicities to that of the Zinn (1985) metallicity
scale. The G4300 and Fe5335 indices are not used for the re-scaling of the
models, since there are fewer GGCs with measurements of these indices. 
 For the magnesium indices, we obtain the linear transformations:

\begin{equation}
\label{eq:mgb}
\rm{[Fe/H]_{z}}=0.631*\rm{[Fe/H]_{Mg~{\it b}}}-0.349, \sigma=0.19~dex\\
\end{equation}

\begin{equation}
\label{eq:mg2}
\rm{[Fe/H]_{z}}=0.799*\rm{[Fe/H]_{Mg_{2}}}-0.230, \sigma=0.16~dex\\
\end{equation}

\noindent for Mg $b$ and Mg$_{2}$ respectively.

The Fe5270 index proved to be non-linear, and was fit with the
quadratic:

\begin{equation}
\label{eq:fe52}
\rm{[Fe/H]_{z}}=-0.186+0.406*\rm{[Fe/H]_{Fe}}-0.393*\rm{[Fe/H]_{Fe}^{2}}\\
\end{equation}

\noindent with an rms of $\sigma$=0.2 dex.

We now derive metallicities from the model predictions for 
our co-added data using the mean\footnote{Strictly speaking,
the indices should be averaged \emph{linearly}, whereas the commonly
used metallicity indicator [Fe/H] is expressed as a logarithmic
scale. However, the difference between the two was only at the level
of $\sim$ 1$\%$, significantly less than the measurement error.}
of the Fe5270, Mg $b$ and Mg$_{2}$ indices, transformed on to the 
Zinn (1985) scale as described above; we do 
not use the Balmer lines in this determination.
We assign weights to each index, corresponding to the inverse of
the scatter from Eqns.~(\ref{eq:mgb} -- \ref{eq:fe52}).
Table~\ref{tab:metallicities} lists 
the metallicities determined for our four mean globular cluster spectra. 
We then compare these metallicities with those predicted by the 
empirical calibrations of Brodie \& Huchra (1990). Their relations 
between the metallicity scale of Zinn (1985) and line-strength 
indices of integrated spectra are based upon observations of globular 
clusters in the Milky Way and M31.
They identify six primary calibrators for metallicity in old stellar
populations, two of which (namely Mg$_{2}$ and Fe5270) we have 
reliably measured  in this study.
 We take the metallicity derived from this calibration as the mean 
predicted by the Mg$_{2}$ and Fe5270 indices measured from 
our spectra, weighted by the reciprocal uncertainty in the index.

\begin{table}
\begin{center}
\caption{\small{Metallicities for our co-added NGC~4472 globular 
clusters using the predictions of the Worthey (1994) models 
(denoted [Fe/H]$_{\rm{W}}$) and the empirical calibrations of 
Brodie \& Huchra (1990) (denoted [Fe/H]$_{\rm{B}}$). 
We also show the metallicities predicted from the $C-T_{1}$ colours using 
Eqn.~(\ref{eq:doug}). Uncertainties in the metallicities are tabulated in
alternate rows.}}
\label{tab:metallicities}
\begin{tabular}{llll}

\hline 

 Bin & [Fe/H]$_{\rm{W}}$ &  [Fe/H]$_{\rm{B}}$ &  
[Fe/H]$_{\rm{GF}}$  \\ 

\hline

1 & -1.29 & -1.3 & -1.34 \\  
... & $\pm$  0.3 & 0.21 & 0.21 \\
2 & -0.91 & -1.03 & -1.00 \\
... & $\pm$  0.35 & 0.31 & 0.16 \\
3 & -0.84 & -0.97 & -0.61 \\
... & $\pm$  0.25 & 0.21 & 0.14\\
4 & -0.27 & 0.30 &  0.10 \\
... & $\pm$  0.30 & 0.30 & 0.26\\ 

\hline

\end{tabular}
\end{center}
\end{table}

The metallicity predictions of the models are generally in good
agreement with the primary calibrators of Brodie \& Huchra (1990).
However, the models predict that our reddest bin is $\sim$ 0.6 dex 
more metal poor than the value  derived from the Brodie \& Huchra (1990)
calibration.
Kissler-Patig \etal (1998) indicate a similar phenomenon for
metal-rich globular clusters in NGC~1399. They find that for Mg$_{2}$ 
$>$ 0.180 mag, the linear extrapolation of Brodie \& Huchra (1990) 
differs noticeably from their metal-rich globular clusters. 
Applying the correction employed by
Kissler-Patig \etal (1998) to our data assigns our 
reddest cluster bin [Fe/H] $\sim$ -0.2, thus bringing the metallicity 
into line with the model predictions. 
The metallicities derived from the flux-weighted Washington colours
using Eqn.~(\ref{eq:doug}) compare well to the
predictions of the Worthey (1994) models, although again they tend to give
 slightly higher metallicities for the redder two bins. 
 
Our cluster bins have a metallicities of approximately -1.6 $\leq$ [Fe/H] 
 $\leq$ 0, which is the range covered by the most metal-rich
two-thirds of the Milky Way globular clusters. Clearly this
is truncated due to the binning of the globular clusters; the actual
colour distribution of the globular clusters from the Geisler \etal (1996) 
catalogue infers a metallicity range from one thousandth to ten times 
solar. At the metal-poor and metal-rich ends, 
this is extrapolated beyond the range covered by GGCs 
from which the colour-[Fe/H] relation was derived, and therefore
carries a greater uncertainty (e.g. Harris \etal 1992).
Our bluest and reddest cluster bins correspond approximately with the 
bimodal peaks of the NGC~4472 globular cluster metallicity distribution. 

The agreement between the metallicities predicted by the SSP models with 
those derived from the empirical calibration and flux-weighted colours is 
encouraging, and gives us some confidence in the model predictions.
Since we will use the SSP models in the age determination of the globular
clusters, we assign the metallicities predicted by the models 
to our globular cluster data as opposed to the mean metallicity 
given by the different methods.

\subsubsection{Radial Gradients in the Globular Cluster Populations}

Geisler \etal (1996) found a radial gradient in the mean C-T$_{1}$ colours 
of the NGC~4472 globular clusters corresponding to 
$\Delta$[Fe/H]/$\Delta$log $r$ $\simeq$ -0.4 dex/log (arcsec), 
$\sigma$ $\sim$ 0.2 dex. 
This they primarily attributed to the varying spatial concentration of the 
blue and red cluster populations, in the sense that the ratio of blue to red
globular clusters increases as a function of galactocentric radius. They also
found that the mean colours of the red globular clusters were similar to that 
of the spheroid light, whilst the colours of the blue population were bluer
than the galaxy light by $\sim$ 0.5 mag at all radii.
 
\begin{figure}
\epsfysize 3.0truein
\hfil{\epsffile{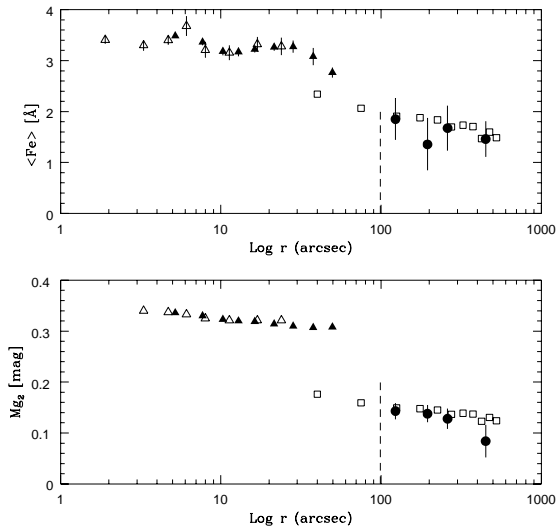}}\hfil
\caption{\small{The behaviour of the $<$Fe$>$ and Mg$_{2}$ indices 
(upper and lower panels respectively) with the logarithm of the projected 
galactocentric radius. Solid circles represent our co-added NGC~4472
globular clusters with their respective bootstrapped uncertainties. 
 The open squares are the predicted index values for 
NGC~4472 globular clusters from the photometry of Geisler \etal (1996)
(see text). We also plot the nuclear data of Davies \etal (1993), both for 
the Cryogenic camera (solid triangles) and the Texas Instruments CCD 
(open triangles).
The dashed line indicates the effective radius of NGC~4472 ($r_{\rm{e}} 
\sim$ 99 arcsec). }}
\label{fig:radial_plot.1}
\end{figure}

We now investigate how the line-strengths of the NGC~4472 globular
clusters behave as a function of projected galactocentric radius.
We first separate our globular cluster data into four radial bins, and 
measure their mean line-strength indices. 
Our spectroscopic sample spans a range of galactocentric radii of 
65 $\le r \le$ 569 arcsec (0.5 $\le r_{\rm{e}} \le$ 6), where
$r_{\rm{e}}$ = 99 arcsec, which is the effective radius of NGC~4472 
(Davies, Sadler \& Peletier 1993). 
To ensure that any detected radial gradient does not originate from a 
varying spatial distribution in the globular cluster subpopulations, 
we fix the ratio of blue:red globular clusters for each bin (we define the 
red globular clusters as those having $C-T_{1}$ $\ge$ 1.625). 
Since the global ratio of
blue to red globular clusters in this study is $\sim$ 2:1, the number of
red globular clusters per bin effectively sets our limiting bin size.
In Fig.~\ref{fig:radial_plot.1} we plot $<$Fe$>$ and Mg$_{2}$ measured for
our data against the logarithm of their mean galactocentric radii.

We also show in Fig.~\ref{fig:radial_plot.1} line-strengths of the
spheroid light of NGC~4472, taken from Davies \etal (1993), who have
 measured the Mg$_{2}$ and $<$Fe$>$ indices out to $\sim$ 0.5 $r_{\rm{e}}$ for
this galaxy. 
To link the radial ranges covered by our data with that of Davies \etal
(1993) more completely, we have converted the $C-T_{1}$ colours of the 
radially binned NGC~4472 globular clusters from Geisler \etal (1996) 
using Eqn.~(\ref{eq:doug}) and the empirical relations of Brodie \& Huchra 
(1990). 
For clarity we have omitted the error bars on the converted photometry, 
but typically the uncertainty in $<$Fe$>$ is $\sim$ 0.4 \AA\ and in Mg$_{2}$
is $\sim$ 0.03 mag.
The conversion from $C-T_{1}$ to $<$Fe$>$ is not a direct one, 
since Brodie \& Huchra (1990) calibrated Fe5270 with metallicity, 
as opposed to the mean of the Fe5270 and Fe5335
indices. We have applied a correction of 0.3 \AA\ to convert the
 Fe5270 prediction to $<$Fe$>$.

The converted globular cluster colour data of Geisler \etal 
(1996) is in good agreement with our $<$Fe$>$ line-strengths at all radii.
A shift of $\sim$ 0.5 dex in $<$Fe$>$ (corresponding to 
removing the blue globular clusters) would make these data overlap with that
of the nuclear line-strength measurements. The metal-rich globular clusters
appear to have similar iron abundances to the spheroid light.
However, the situation for Mg$_{2}$ appears somewhat different.
Again, our data are consistent with the indices predicted by the 
globular cluster photometry of Geisler \etal (1996).
However, the GCs are offset downwards from the spheroid data of 
Davies \etal (1993) by $\sim$ -0.10 mag at the same
radii. 
Assuming the data of Geisler \etal (1996) traces the behaviour of the 
GCs inwards to 0.5 r$_{e}$, then even after applying an additive shift
of 0.05 mag (for the metal-rich GCs), there is still a 
difference of $\sim$ 0.05 mag between the GCs and spheroid light. 

It is evident that there is a metallicity gradient in our co-added data, 
in the sense that Mg$_{2}$ and $<$Fe$>$ weaken with increasing 
galactocentric radius. 
We see no significant trend of changing flux in the spectra with radius, 
which could possibly introduce an artificial radial gradient.
Applying a weighted linear fit to the $<$Fe$>$ and Mg$_{2}$
indices, we obtain $\Delta$[Fe/H]/$\Delta$ log $r$ = -0.30 $\pm$ 0.21 dex 
for  $<$Fe$>$
and $\Delta$[Fe/H]/$\Delta$ log $r$ = -0.94  $\pm$ 0.31 for Mg$_{2}$. 
The steep gradient
seen in Mg$_{2}$ is primarily driven by the outermost radial bin (which 
has the largest uncertainties). Removing this
point from the fit yields a gradient of $\Delta$[Fe/H]/$\Delta$ log $r$ = 
-0.45  $\pm$ 0.19 dex. 
Both of these values are consistent with the mean gradient found 
by Geisler \etal (1996) of $\Delta$[Fe/H]/$\Delta$ log $r$ = 
-0.4 $\pm$ 0.2 dex, and the mean gradient found for
GCS's in general of $\Delta$[Fe/H]/$\Delta$ log $r$ = -0.5 dex 
(Ashman \& Zepf 1998).
By comparison, the gradient from the spheroid data of Davies 
\etal (1993) is $\Delta$[Fe/H]/$\Delta$ log $r$ = -0.20 $\pm$ 0.10 dex, 
similar to the value found by Kim \etal(2000) using Washington
photometry (for $r \le$ 180 arcsec).

\begin{figure}
\epsfysize 3.0truein
\hfil{\epsffile{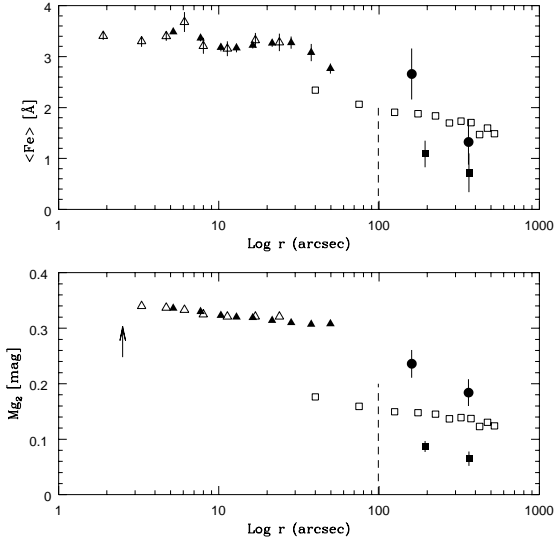}}\hfil
\caption{\small{The behaviour of the $<$Fe$>$ and Mg$_{2}$ indices with the 
logarithm of the galactocentric radius for the two subpopulations of
globular clusters. Solid circles represent the red 
($C-T_{1} \ge$ 1.625) globular clusters, solid squares represent the 
blue globular clusters.
The triangles denote the nuclear data of Davies \etal (1993), both for the
Cryogenic camera (solid) and the Texas Instruments CCD (open triangles).
The dashed line indicates the effective radius of NGC~4472.
The arrow in the lower panel shows the maximum corrective vector for [Mg/Fe] 
overabundance taken from Henry \& Worthey (1999).
}}
\label{fig:radial_plot.2}
\end{figure}

To investigate this matter further, and to directly compare the spheroid light
with the globular cluster subpopulations, we have binned our globular 
clusters into a blue and red population, taking the colour cut to be 
$C-T_{1}$ = 1.625. 
We have further split each of these two bins by radius into two equal 
components and show our results in Fig.~\ref{fig:radial_plot.2}.

The Mg$_{2}$ and $<$Fe$>$ indices decrease with increasing galactocentric 
radius for \emph{both} the red and blue globular cluster subpopulations.
Formally, the radial gradients of the clusters from spectroscopy
are twice as steep as those determined from the photometric data.
However, the significance of this, particularly for the red globular
clusters (for which \emph{n} $\sim$ 23) is marginal.
Measured in $<$Fe$>$, the innermost red cluster bin is 
 comparable to the spheroid light from Davies \etal (1993), 
whereas the Mg$_{2}$ index of this bin is some 0.05 mag lower. 
In the lower panel of Fig.~\ref{fig:radial_plot.2}, we plot a vector
corresponding to the maximum correction expected for [Mg/Fe] 
overabundant ratios seen in the brightest elliptical galaxies, 
 of which NGC~4472 is known to be affected (Worthey \etal 1992; 
Henry \& Worthey 1999; Kobayashi \& Arimoto 1999).
If we apply this maximum correction ($\sim$0.05 mag) to the Mg$_{2}$ 
index of the GCs, then these data become consistent with the 
spheroid data of Davies \etal (1993). 
Since the $<$Fe$>$ index of the globular clusters is consistent with 
that of the spheroid light with no correction, we tentatively 
conclude that the globular clusters in NGC~4472 show no sign of non-solar 
[Mg/Fe] ratios. We return to this point in $\S 5.4$.

\subsection{Cluster Ages}

We now turn to the problem of deriving mean ages for our globular
clusters using the Worthey (1994) models.
Our primary age indicator for these data is the H$\beta$ index, 
therefore the uncertainties in this index will reflect heavily 
on our age estimates for the globular clusters. 
We calibrate the models to the canonical age we have chosen for the
GGCs, which is 12 Gyr.
This age for the GGCs is consistent with recent work using 
the latest \textsc{hipparcos} parallaxes by Carretta \etal (2000), 
and is convenient since this age-line is directly modelled 
(i.e. not interpolated) by Worthey (1994). 
We fit the 12 Gyr line of the models to the H$\beta$ index of the GGCs 
using a $\chi^{2}$ procedure.
The shift implied in H$\beta$ is then applied to our NGC~4472 globular
cluster data. This correction to the models is typically $\sim$ -0.4 \AA, 
larger than modelling uncertainties, and is effectively a zero-point
calibration. Without this shift, nearly \emph{all} the GGCs would be 
predicted to have ages $\ge$ 18 Gyr.

\begin{figure}
\epsfysize 3.0truein
\hfil{\epsffile{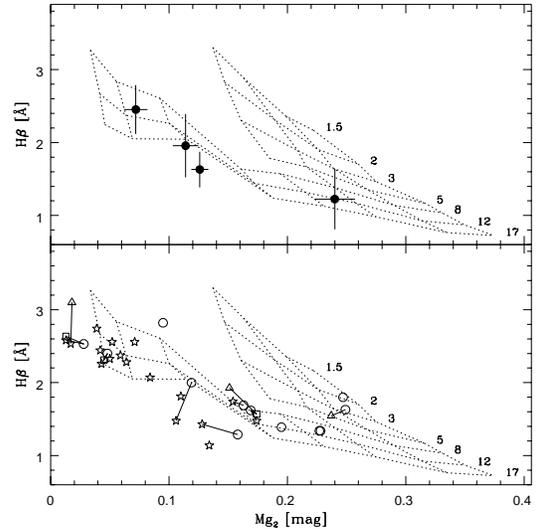}}\hfil
\caption{\small{Predictions of the Worthey (1994) models
for these data. Top panel: our co-added NGC~4472 GCs are represented
by filled circles with error bars. 
 Lower panel: data for the combined sample of GGCs. 
Open circles are GGCs from Cohen \etal (1998), 
open stars are data from Brodie \& Huchra (1990), triangles are
our CFHT data and squares are our WHT data. Lines connect the same
GGCs between different datasets.
 The scale on the right side of the grid gives 
age predictions of the models ranging from 1.5 -- 17 Gyr.
}}  
\label{fig:wor_mg2_hbeta}
\end{figure}

\begin{figure}
\epsfysize 3.0truein
\hfil{\epsffile{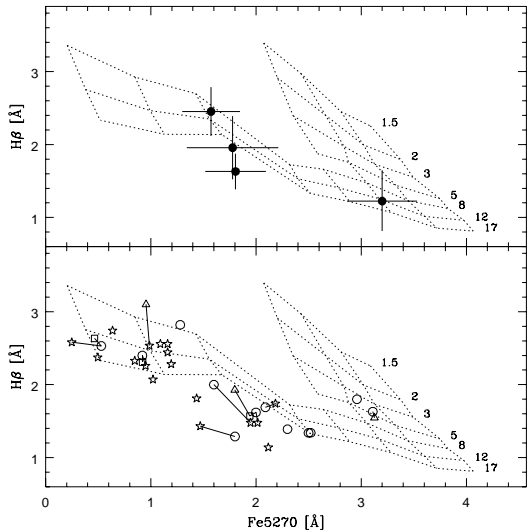}}\hfil
\caption{\small{H$\beta$ index against Fe5270. 
Symbols are the same as for Fig~\ref{fig:wor_mg2_hbeta}.}}
\label{fig:wor_fe52_mgb}
\end{figure}

\begin{figure}
\epsfysize 3.0truein
\hfil{\epsffile{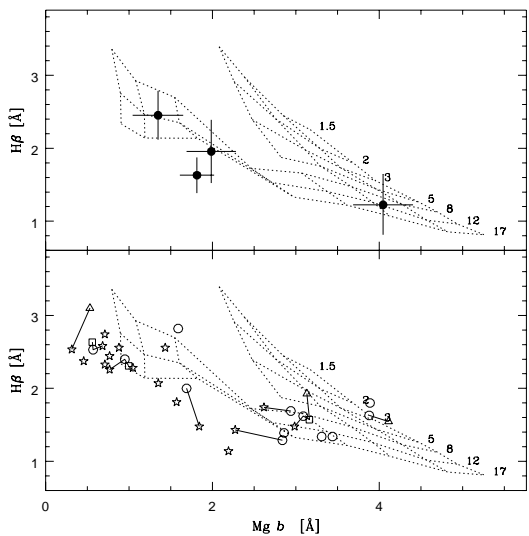}}\hfil
\caption{\small{H$\beta$ index versus Mg $b$. 
Symbols are the same as for Fig~\ref{fig:wor_mg2_hbeta}.}}
\label{fig:wor_mgb_hbeta}
\end{figure}

\begin{figure}
\epsfysize 3.0truein
\hfil{\epsffile{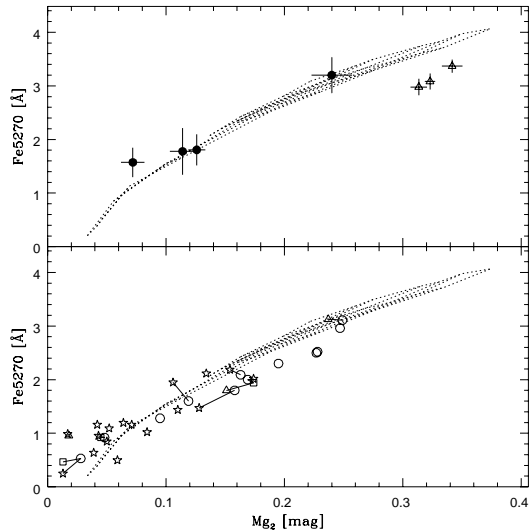}}\hfil
\caption{\small{Fe5270 index versus Mg$_{2}$. 
Symbols same as Fig~\ref{fig:wor_mg2_hbeta}, but also including the
spheroid data of Davies \etal (1993) which we have binned in radius
(open triangles, top panel). 
The age scale of the models has been omitted for clarity.}} 
\label{fig:wor_mg2_fe52}
\end{figure}

We compare our data to the corrected models in 
Figs.~(\ref{fig:wor_mg2_hbeta} -- \ref{fig:wor_mg2_fe52}), 
each of which consist of two panels: the upper panel over-plots the SSP 
grids of Worthey (1994) on to our co-added data (filled circles with
bootstrapped uncertainties). The lower panel compares the combined
sample of GGCs to these same models.

\begin{figure}
\epsfysize 3.0truein
\hfil{\epsffile{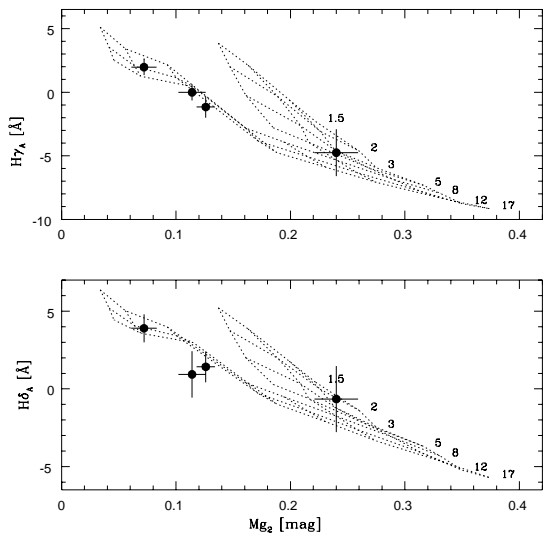}}\hfil
\caption{\small{The Worthey (1994) model predictions for our two
other age indicators. Top panel: H$\gamma_{\rm{A}}$ versus Mg$_{2}$, bottom
panel: H$\delta_{\rm{A}}$ versus Mg$_{2}$.}}
\label{fig:balmer}
\end{figure}

On the whole, the models give a good reproduction of the behaviour of
our data, both for the co-added NGC~4472 globular clusters and 
the GGCs. The Balmer indices track the isochrones of the models
fairly closely, across two decades in metallicity.
The new modelling of the cluster HBs shows strikingly similar
behaviour to that of our three most metal-poor cluster bins.
It is apparent that there are small shifts between different 
metal indices, but these are typically 0.2 \AA\ and well
within the uncertainties. 

In Fig.~\ref{fig:wor_mg2_fe52} we plot Fe5270 against
Mg$_{2}$ for our data and the combined sample of GGCs.
For comparison, we also plot the nuclear data of Davies \etal (1993), 
which we have placed into three radial bins. 
The spheroid data of Davies \etal (1993)
is clearly offset from the models, which is usually interpreted
as an Mg$_{2}$ overabundance (Davies \etal 1993).
Unlike the stellar spheroid light, the NGC~4472 globular clusters show no 
evidence of Mg$_{2}$ enhancement over iron. This is consistent with
the discussion of radial gradients in $\S$ 5.3.2

We have already derived metallicities for the GGCs
and NGC~4472 globular clusters, and effectively calculate age
as a function of this metallicity since the H$\beta$ (age) 
lines of the models are calculated with respect to the 
metal indices. Ages for the NGC~4472 globular clusters are 
determined using the isochrones of the Worthey (1994) models 
predicted for the Fe5270, Mg$_{2}$, Mg $b$ and H$\beta$ indices along with 
their respective uncertainties.
Our final ages for the co-added globular clusters are the mean age derived from
these indices, and are given in Table~\ref{tab:ages} along with
our final metallicities for each bin.
In view of the large formal uncertainties, it is clear from 
Table~\ref{tab:ages} that there is no evidence of a 
substantially younger population of globular clusters present in our
data. The most metal-rich and 
metal poor bins appear to be younger than the two intermediate ones. 
This is likely an artefact of the Worthey (1994) models at metallicities 
-1.0 $<$ [Fe/H] $<$ -0.5, the regime were HB morphology starts to play 
a role.
The switch to M3 morphology happens below [Fe/H] = -0.95 and the switch 
to M92 morphology happens below [Fe/H] =-1.55. This change does not occur
as a smoothly varying function of metallicity, and therefore may be
predicting too sharp a rise in the H$\beta$ index. 
The models also predict too old an age for the GGCs at these metallicities. 
Since the 12 Gyr age isochrone of the models is fit to the GGCs, 
this may have the effect of inducing too large a shift
in H$\beta$ and therefore predicting younger ages in the co-added data.
Formally, the mean age of the four bins is 14 $\pm$ 6 Gyr, consistent with our
canonical age adopted for the GGCs of 12 Gyr.

\begin{table}
\begin{center}
\caption{\small{The final ages and metallicities for our co-added data, 
derived from the stellar population models of Worthey (1994). The
 final ages in column 3 are the mean of those predicted by the 
H$\beta$ index as a function of Fe5270, Mg $b$ and Mg$_{2}$.}}
\label{tab:ages}
\begin{tabular}{ccc}

\hline 

 Bin & [Fe/H]$_{\rm{Z}}$ & Mean Age \\
 \# & (dex) & (Gyr)\\  

\hline
1 & -1.29 $\pm$ 0.30 & 10.7 $^{+6}_{-5}$\\  
2 & -0.91 $\pm$ 0.35 & 15.3 $^{+8}_{-5}$\\  
3 & -0.84 $\pm$ 0.25 & 18.5 $^{+4}_{-5}$\\  
4 & -0.27 $\pm$ 0.30 & 11.3 $^{+8}_{-9}$\\  
\hline

\end{tabular}
\end{center}
\end{table}

We have also measured two other primarily age-sensitive indices, namely
H$\gamma_{\rm{A}}$ and H$\delta_{\rm{A}}$, which we show in 
Fig.~\ref{fig:balmer}.
Both these indices exhibit similar behaviour to H$\beta$ measured for the
globular clusters, but with indications that the most metal-rich bin is
somewhat younger than the others. However, not only is the correction of the 
models to these data more uncertain (since we have few GGCs for which there
are H$\gamma_{\rm{A}}$ and H$\delta_{\rm{A}}$ measurements), 
but also the uncertainties 
in these indices are somewhat larger, and therefore we do not derive ages for
the globular clusters from these higher-order Balmer lines.

\begin{figure}
\epsfysize 3.0truein
\hfil{\epsffile{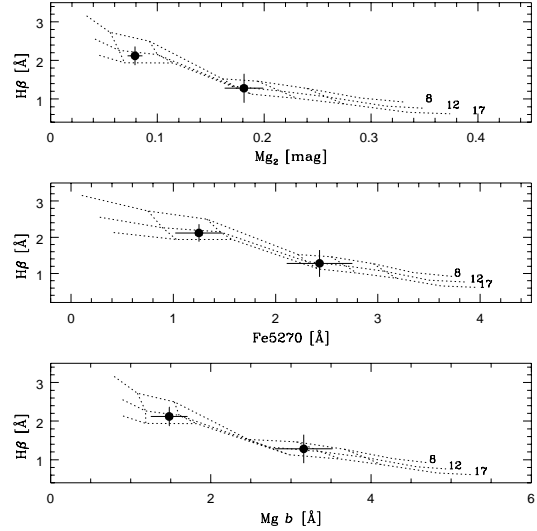}}\hfil
\caption{\small{Age predictions of the Worthey (1994) models
for the H$\beta$ index versus Mg$_{2}$, Fe5270 and Mg $b$ for 
our metal-poor and metal-rich bins of NGC~4472 globular clusters. 
We show only the $\tau \geq$ 8 Gyr isochrones for clarity. 
}}
\label{fig:panels}
\end{figure}

In order to reduce the uncertainties in the indices (especially
H$\beta$), but at the expense of losing our baseline in metallicity, 
we have also binned our sample of NGC~4472 globular clusters into 
one metal-poor globular cluster population and one metal-rich population. 
We take the dividing line between the populations as $C-T_1$ = 1.625, 
obtained from the KMM mixture-modelling test (Ashman \etal 1994) applied
to the full NGC~4472 globular cluster data of Geisler \etal (1996).
Again we co-add these data and measure the Lick/IDS indices using the 
bootstrap procedure described previously. 
Our metal-poor and metal-rich bins contain 85 and 45
spectra respectively, with flux weighted colours of  $C-T_1$ = 1.4 and
 $C-T_1$ = 1.8, corresponding to metallicities of [Fe/H] = -1.10 
and [Fe/H] = -0.16 using Eqn.~(\ref{eq:doug}).
We plot Mg $b$, Fe5270 and Mg$_{2}$ against 
H$\beta$ for these bins in Fig.~\ref{fig:panels}.
Again, the models have been normalised to the canonical ages adopted for
the GGCs. 
From these indices, we obtain a mean age of 14.5 $\pm$ 4 Gyr for
the metal-poor bin and 13.8 $\pm$ 6 Gyr for the metal-rich bin. 
The smaller uncertainties in the metal-poor bin correspond to
the larger numbers of blue globular cluster spectra available for co-addition
(for 40 $\leq n \leq$ 100 the uncertainty in the
index goes as $\sigma \propto n^{-0.3}$). 

As a check for systematic errors in our age results, we perform Monte Carlo
simulations on the Worthey (1994) model SEDs predicted for the ages and 
metallicities of our metal-poor and metal-rich globular cluster bins.
We obtain SEDs corresponding to the indices measured for each bin and 
then artificially degrade them following a procedure similar to
that described in $\S$ 4.2. 
We generate the same number of realisations as there are spectra in
each of our red and blue bins.
We then measure the ages of the co-added artificial spectra, again using the 
Worthey (1994) grids.
We find that each of our cluster bins is older than 6 Gyr at the 
95 \% confidence level, and older than 3 Gyr at the 98.5 \% confidence 
level.

From HST $V$ and $I$ data, Puzia \etal (1999) find that the red and blue 
GCs in NGC~4472 are old and coeval within their errors, but that formally the 
\emph{blue} GCs are somewhat younger. 
Lee \& Kim (2000) conclude that the \emph{red} GCs are
younger than the blue by 'several' Gyr from the same data (plus one additional
HST field), attributing their differences to a larger sample size, increased 
areal coverage and greater completeness.

Our findings that the blue and red globular clusters of NGC~4472 are old 
and coeval within the uncertainties (but formally the red GCs are younger) 
are not inconsistent with either of the above two studies, 
but do not rely on any assumption about the underlying mass distributions of 
the globular cluster populations.

\section{Discussion}

The major conclusions of this work are that the NGC~4472 globular 
cluster subpopulations are essentially coeval and old, and 
separated by $\sim$ 1.0 dex in [Fe/H]. The NGC~4472 GCS exhibits a radial
metallicity gradient steeper than the underlying halo
light of the galaxy, with the red and blue globular clusters both showing
some evidence of this behaviour. 
The globular cluster populations of NGC~4472 clearly formed from gas with 
significantly different levels (perhaps histories) of enrichment, 
but at roughly similar times.
A strong radial metallicity gradient in the globular clusters 
would imply a largely dissipative formation mechanism, which classically 
points toward some form of monolithic protogalactic collapse
(e.g. Eggen, Lynden-Bell \& Sandage 1962), or modified 
multiphase collapse model (e.g. Forbes, Brodie \& Grillmair 1997).
A gradient of $\Delta$[Fe/H]/$\Delta$ log $r$ $\simeq$ -0.4 for the GCs
is well within the predictions of theoretical models of the formation
of spheroids. For example, the dissipative models of Larson (1974) predict 
$\Delta$[Fe/H]/$\Delta$ log $r$ $=$ -1.0, whilst the variants of 
Carlberg (1984) range from -0.5 to 0.0.
However, these models lead to large rotation at high metallicity, which
is not observed for this galaxy (e.g. Zepf \etal in preparation).
Large radial gradients also do not preclude the possibility that major
mergers may have contributed to the formation of NGC~4472 and its GCS
(e.g. Ashman \& Zepf 1992). As simulations show 
(e.g. White 1980; Barnes 1988), a radial gradient is not necessarily 
destroyed during the merger of disks, but may simply become somewhat diluted.

In terms of abundance ratios, the variation seen in [Mg/Fe] is generally 
attributed to different levels of enrichment from Type Ia 
(largely contributing Fe) and Type II (Mg and Fe with all other elements) 
supernovae. If the NGC~4472 globular clusters truly possess [Mg/Fe] $\simeq$ 
0, then this would imply that their formation was decoupled (at some level)
from that of the spheroid light. 
Whether differing degrees of Type Ia/Type II enrichment 
indicates different star formation time-scales, a variable binary fraction
or a variable IMF is uncertain.  
But certainly determining whether the [$\alpha$/Fe] ratios in 
globular clusters are truly enhanced with respect to solar, and whether 
there are variations \emph{between} different cluster
subpopulations will place strong constraints on their formation.

Measuring age differences between old stellar populations is
notoriously difficult. Indeed if the metal-poor and metal-rich 
globular clusters really \emph{are} separated in age by several Gyr, 
how may we proceed to detect this difference? The automatic response
from this work would be to state that better quality, high S/N spectra of
individual globular clusters are required for accurate age-dating.
One step would be to move away from instrument-specific systems which often 
require many correction factors and the degradation in resolution of ones 
data. This is effectively throwing away information, a better route perhaps
is followed by Vazdekis \etal (1996) whereby the models are freely available
to be adapted to the characteristics of these data, in order to maximize 
the information obtained from these data.
Globular clusters are essentially homogeneous and coeval associations of stars,
and therefore the assumptions implicit in SSP models are well suited to 
these systems. The fact that observed ages and metallicities are 
luminosity--weighted is less significant for globular clusters, unlike the 
situation for composite stellar systems (i.e. galaxies) whose integrated light 
may be dominated by the most recent star formation event.
It is unfortunate then, that the sub-solar regime of the models (the 
parameter space of the majority of globular clusters) is poorly constrained.
At the most metal-poor end, SSP model grids are uncertain by many Gyr for a
 given measurement of H$\beta$. This largely stems from the lack
of metal-poor stars in the solar neighbourhood, and the assumptions 
made for HB morphology. Moreover, the SSP models
are based upon stellar libraries of stars of solar abundance ratios,
whereas the abundances of the most luminous ellipticals are clearly 
non-solar (e.g. Worthey \etal 1992). Abundances obtained from
$\alpha$-elements are typically much higher than those derived from the 
iron-peak, and this is the principle reason for uncertainty in the absolute
abundance and age scales.

\begin{figure}
\epsfysize 3.0truein
\hfil{\epsffile{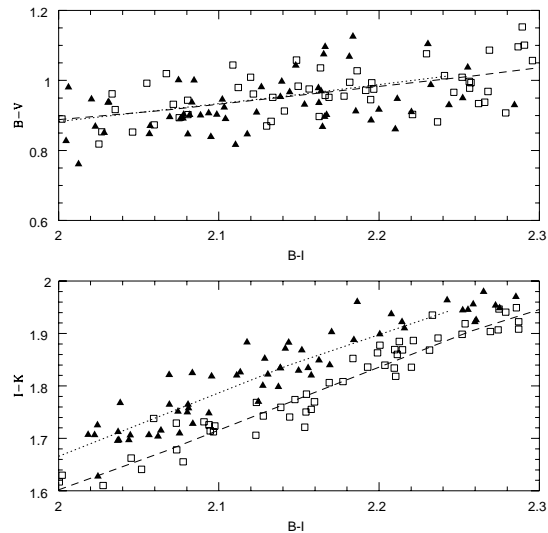}}\hfil
\caption{\small{The predicted $B$, $V$, $I$ and $K$ colours for a subset of
NGC~4472 globular clusters (Geisler \etal 1996) using the stellar
population models of Vazdekis (in preparation). 
Solid triangles represent the red ($C-T_{1} \ge$1.625) globular clusters, 
with an assumed age of 10 Gyr, open squares represent the blue globular
clusters, with an assumed age of 15 Gyr. We have included a representative
observational scatter of 0.05 mag in the colours.
The top panel highlights the degeneracy between age and metallicity in
optical broadband colours, whereas the inclusion of the $K$-band (lower panel)
can distinguish between the two cluster populations. Dashed and dotted lines
indicate the 15 Gyr and 10 Gyr isochrones of the models respectively.}} 
\label{fig:colours}
\end{figure}

Spectroscopy is the only way of unambiguously identifying 
\emph{bona fide} globular clusters (through their kinematics),
in addition to investigating $\alpha$-element enhancement. 
However, building up statistically significant samples of high S/N
spectra for globular clusters is time consuming, even on
8-metre class telescopes.
An alternative route towards disentangling the ages and metallicities for
globular clusters is through the use of two colour
photometry, with the inclusion of an infrared passband
(Kissler-Patig 2000).
SSP models of broadband optical colours are effectively
degenerate, and isochrones in these models ranging from 1 -- 17 Gyr 
are largely superposed upon lines of constant metallicity. 
However, a colour such as $I-K$ directly measures the temperature
of the red giant branch in old stellar populations, which is
dependent almost entirely on metallicity. By plotting a 
metallicity sensitive colour against an age/metallicity sensitive
colour allows the extraction of age information (albeit still requiring 
high photometric accuracy, $\simeq$ 0.05 mag).
Fig~\ref{fig:colours} illustrates the effect of including the
$K$-band in two-colour SSP model predictions.
We have taken a subset of $\sim$ 300 globular clusters from the 
catalogue of Geisler \etal (1996), and have assigned the
blue globular clusters an age of 15 Gyr, and the red 
($C-T_{1}$ $\ge$ 1.625) globular clusters an age of 10 Gyr.
We then plot the predictions of the SSP models of Vazdekis (in 
preparation) for the $B$, $V$, $I$ and $K$ bands at these ages and 
metallicities, including an observational uncertainty of 0.05 mag in 
the colours.
The optical colours are clearly degenerate, but the inclusion of
the $K$-band provides a powerful discriminant between age and
metallicity. 

\section{Summary}

We have derived ages and metallicities from 130 co-added globular cluster
spectra associated with the giant elliptical NGC~4472. 
We have measured metallicity sensitive and age sensitive line indices from
the spectra and compared them to the stellar population models of 
Worthey (1994).
We find that the globular clusters span a metallicity range of 
approximately -1.6 $\leq$ [Fe/H] $\leq$ 0 dex, corresponding to the range 
covered by the more metal-rich two-thirds of the Milky Way globular cluster 
system.
Although absolute ages are uncertain, via the calibration of 
the stellar population models with Galactic Globular Clusters, we find that 
the metal-poor population of NGC~4472 globular clusters has an age of
14.5 $\pm$ 4 Gyr, whilst the metal-rich population is 13.8 $\pm$ 6 Gyr old.
Monte Carlo simulations indicate that both the globular cluster populations 
present in these data are older than $\tau =$ 6 Gyr at the 95 \% 
confidence level.
The NGC~4472 globular cluster system exhibits a radial metallicity gradient 
steeper than the underlying spheroid light of the galaxy even after 
correction for the varying spatial distribution of the blue and red globular 
cluster populations.
We see no evidence for non-solar [Mg/Fe] ratios for the higher metallicity 
globular clusters, at the level observed for the spheroid light.
This would indicate that the formation of the NGC~4472 globular clusters was 
probably decoupled from that of the spheroid stars.

\section{Acknowledgements}

MB would like to acknowledge Guy Worthey and Alexandre Vazdekis for the use 
of their stellar population models, and PPARC for its supporting studentship. 
Thanks also go to Harald Kuntschner and John Blakeslee, who both provided 
useful comments regarding this paper, and the referee, Raffaelle Gratton, 
for his perceptive reading of the paper.
MB also acknowledges the use of the STARLINK facilities at the
University of Durham.

\label{lastpage}

\newpage
\begin{thebibliography}{}

\bibitem[\protect\citename{Aaronson}1978]{}Aaronson M., Cohen J.G., Mould J., Malkan M., 1978, ApJ, 223, 824
\bibitem[\protect\citename{Armandroff}1988]{}Armandroff T.E., Zinn R.J., 1988, AJ, 96, 92
\bibitem[\protect\citename{Ashman}1992]{}Ashman K.M., Zepf S.E., 1992, ApJ, 384, 50
\bibitem[\protect\citename{Ashman}1994]{}Ashman K.M., Bird C.M., Zepf S.E.,
1994, 108, 2348
\bibitem[\protect\citename{Ashman1}1998]{}Ashman K.M., Zepf S.E., 1998, Globular Cluster Systems, eds., King A., Lin D., Maran S., Pringle J., Ward M., Cambridge University Press
\bibitem[\protect\citename{Barnes}1988]{}Barnes J.E., 1988, ApJ, 331, 699
\bibitem[\protect\citename{Brodie}1990]{}Brodie J.P., Huchra J.P., 1990, ApJ, 362, 503
\bibitem[\protect\citename{Brodie}1991]{}Brodie J.P., Huchra J.P., 1991, ApJ, 379, 157
\bibitem[\protect\citename{Burstein}1984]{}Burstein D., Faber S.M., Gaskell C.M., Krumm N., 1984, ApJ, 287, 586
\bibitem[\protect\citename{Carlberg}1984]{}Carlberg R.G., 1984, ApJ, 286, 403
\bibitem[\protect\citename{Carretta}1997]{}Carretta E., Gratton R.G., 1997, A\&AS, 121, 95
\bibitem[\protect\citename{Carretta}2000]{}Carretta E., Gratton R.G., Clementini G., Fusi-Pecci F., 2000, ApJ, 533, 215
\bibitem[\protect\citename{Cohen1}1988]{}Cohen J.G., 1988, AJ, 95, 982
\bibitem[\protect\citename{Cohen}1998]{}Cohen J.G., Blakeslee J.P., Ryzhov A., 1998, AJ, 496, 808
\bibitem[\protect\citename{Couture}1991]{}Couture J., Harris W.E., Allwright J.W.B., 1991, ApJ, 372, 97
\bibitem[\protect\citename{Davies}1993]{}Davies R.L., Sadler E.M., Peletier R.F., 1993, MNRAS, 262, 650
\bibitem[\protect\citename{Eggen}1962]{}Eggen O.J., Lynden-Bell D., Sandage A.R., 1962, ApJ, 136, 748
\bibitem[\protect\citename{Faber}1972]{}Faber S.M., 1972, A\&A, 20, 361
\bibitem[\protect\citename{Faber}1985]{}Faber S.M., Friel E. D., Burstein D., Gaskell C. M., 1985, ApJS, 57, 711
\bibitem[\protect\citename{Forbes}1997]{}Forbes D.A., Brodie J.P., Grillmair C.J., 1997, AJ, 133, 1652
\bibitem[\protect\citename{Gebhardt}1999]{}Gebhardt K., Kissler-Patig M., 1999, AJ, 
118, 1526
\bibitem[\protect\citename{Geisler}1990]{}Geisler D., Forte J.C., 1990, ApJ, 350, 5
\bibitem[\protect\citename{Geisler1}1996]{}Geisler D., Lee M.G., Kim E., 1996, AJ, 111, 1529
\bibitem[\protect\citename{Gonzalez}1993]{}Gonzal$\acute{e}$z J.J., 1993, PhD thesis, University of California, Santa Cruz
\bibitem[\protect\citename{Gorgas}1993]{}Gorgas J., Faber S.M., Burstein D., Gonzal$\acute{e}$z J.J., Courteau S., Prosser C., 1993, ApJS, 86, 153
\bibitem[\protect\citename{Hanes}1986]{}Hanes D.A., Brodie J.P., 1986, ApJ, 300, 279
\bibitem[\protect\citename{Harris}1977]{}Harris H.C., Canterna R., 1977, AJ, 82, 798 
\bibitem[\protect\citename{Harris}1992]{}Harris G.L.H., Geisler D., Harris H.C., Hesser J.E., 1992, AJ, 104, 613
\bibitem[\protect\citename{Henry}1999]{}Henry R.B.C., Worthey G., 1999, PASP, 111, 919
\bibitem[\protect\citename{Kim}2000]{}Kim E., Lee M.G., Geisler D., 2000, in press, astro-ph/0001007
\bibitem[\protect\citename{KisslerPatig}2000]{}Kissler-Patig M., 2000, astro-ph/0002070
\bibitem[\protect\citename{KisslerPatig}1998]{}Kissler-Patig M., Brodie J.P.,
Schroder L.L., Forbes D.A., Grillmair C.J., Huchra J.P., 1998, AJ, 115, 105
\bibitem[\protect\citename{Kobayashi}1999]{}Kobayashi C., Arimoto N., 1999, AJ, 527, 573
\bibitem[\protect\citename{Kundu}1999]{}Kundu A., Whitmore B.C., Sparks W.B., Macchetto F.B., Zepf S.E., Ashman K.E., 1999, ApJ, 513, 733 
\bibitem[\protect\citename{Larson}1974]{}Larson R.B., 1974, MNRAS, 166, 585
\bibitem[\protect\citename{Lee1}2000]{}Lee M.G., Kim E, 2000, astro-ph/0004116
\bibitem[\protect\citename{Lee}1998]{}Lee M.G., Kim E, Geisler D., 1998, AJ, 115, 947
\bibitem[\protect\citename{Meylan}1997]{}Meylan G., Heggie D.C., 1997,  Astron. Astrophys. Rev., 8, 1
\bibitem[\protect\citename{Mould}1990]{}Mould J.R., Oke J.B., de Zeeuw P.T., Nemec J.M., 1990, AJ, 99, 1823
\bibitem[\protect\citename{Neilsen}1999]{}Neilsen E.H., Tsvetanov Z.I., 1999, ApJ, 515, 13
\bibitem[\protect\citename{O'Connell}1976]{}O'Connell R.W., 1976, ApJ, 206, 370
\bibitem[\protect\citename{Puzia}1999]{}Puzia T.H., Kissler-Patig M., Brodie J.P., Huchra J.P., 1999, AJ, 118 ,2734
\bibitem[\protect\citename{Sandage}1981]{}Sandage A., Tammann G.A., 1981, A Revised Shapley-Ames Catalogue of Bright Galaxies (Washington : Carnegie Inst. Washington)
\bibitem[\protect\citename{Sharples}1998]{}Sharples R.M., Zepf S.E., 
Bridges T.J., Hanes D.A., Carter D., Ashman K.M., Geisler D., 1998, AJ, 115, 2337
\bibitem[\protect\citename{Tonry}1979]{}Tonry J., Davis M., 1979, AJ, 84, 151
\bibitem[\protect\citename{Vazdekis}1996]{}Vazdekis A., Casuso E., Peletier R.F., Beckman J.E., 1996, ApJS, 106, 307
\bibitem[\protect\citename{White}1980]{}White S.D.M., 1980, MNRAS, 191, 1
\bibitem[\protect\citename{Worthey4}1992]{}Worthey G., Faber S.M., Gonzal$\acute{e}$z, J.J., 1992, AJ, 398, 69
\bibitem[\protect\citename{Worthey}1993]{}Worthey G., 1993, ApJ Lett., 415, L91
\bibitem[\protect\citename{Worthey1}1994]{}Worthey G., 1994, ApJS, 95, 107 
\bibitem[\protect\citename{Worthey2}1997]{}Worthey G., Ottaviani D.L., 1997, ApJS, 111, 377
\bibitem[\protect\citename{Worthey1}1994]{}Worthey G., Faber S.M., Gonzal$\acute{e}$z J.J., Burstein D., 1994, ApJS, 94, 687

\bibitem[\protect\citename{Zepf}1993]{}Zepf S.E., Ashman K.M., 1993, MNRAS, 264, 611 
\bibitem[\protect\citename{Zepf1}1999]{}Zepf S.E., Ashman K.M., English J., Freeman K.C., Sharples R.M. 1999, AJ, 118, 752
\bibitem[\protect\citename{Zinn1}1985]{}Zinn R. 1985, ApJ, 293, 424 

\end{thebibliography}
\end{document}